\documentclass[conference, square, comma]{IEEEtran}

\usepackage{amsmath,amssymb,amsfonts}
\usepackage{graphicx}
\usepackage{textcomp}
\usepackage{xcolor}
\usepackage{hyperref}

\usepackage{amsthm}
\newtheorem{definition}{Definition}
\usepackage{mathrsfs}
\def\BibTeX{{\rm B\kern-.05em{\sc i\kern-.025em b}\kern-.08em
    T\kern-.1667em\lower.7ex\hbox{E}\kern-.125emX}}

\usepackage{algorithm}
\usepackage{algorithmicx}
\usepackage[noend]{algpseudocode}
\usepackage{multirow} 
\usepackage{booktabs}
\usepackage{tabularray}
\usepackage[export]{adjustbox}
\usepackage{times}
\usepackage{xcolor}
\usepackage{tablefootnote}
\usepackage[square,numbers]{natbib}
\setcitestyle{sort&compress}
\usepackage{xspace} % for \xspace
\newcommand{\pks}[1]{#1}
\newcommand{\pk}[1]{#1}
\newcommand{\name}{\textsc{Shrink}\xspace} %change name any time
\newcommand{\ggd}{\textsc{GreedyGD}\xspace} %change name any time
\newcommand{\simpiece}{\textsc{SimPiece}\xspace} %change name any time
\newcommand{\ourname}{SHRINK\xspace} %change name any time
\newcommand{\origin}{\Theta} %change name any time
\newcommand{\Span}{\Psi} %change name any time
\newcommand{\semantic}{\mathscr{S}}

\algrenewcommand{\algorithmiccomment}[1]{\textit{\hfill// #1}}

\renewcommand{\footnoterule}{
    \kern -3pt
    \hrule width 0.95\columnwidth
    \kern 2.6pt
}

\begin{document}

\title{\name: Data Compression by Semantic Extraction and Residuals Encoding}
% \title{\name Data with Semantic Strengthening and Residual Encoding}

\author{
\IEEEauthorblockN{Guoyou Sun\textsuperscript{a}, Panagiotis Karras\textsuperscript{b,c}, Qi Zhang\textsuperscript{a}}
\IEEEauthorblockA{
 \textit{\textsuperscript{a} DIGIT and Department of Electrical and Computer Engineering, Aarhus University, Denmark} \\
 \textit{\textsuperscript{b} DIGIT and Department of Computer Science, Aarhus University, Denmark} \\
 \textit{\textsuperscript{c} Department of Computer Science, University of Copenhagen, Denmark} \\
Email: \{guoyous, qz\}@ece.au.dk,  piekarras@gmail.com
}
}

% \author{\IEEEauthorblockN{Guoyou Sun}
% \IEEEauthorblockA{
% % \textit{dept. name of organization (of Aff.)} \\
% \textit{Aarhus University}\\
% % Aarhus, Denmark \\
% guoyous@ece.au.dk
% }
% \and
% \IEEEauthorblockN{Panagiotis Karras}
% \IEEEauthorblockA{
% % \textit{dept. name of organization (of Aff.)} \\
% \textit{U. of Copenhagen \& Aarhus U.}\\
% % Aarhus, Denmark \\
% piekarras@gmail.com}
% \and
% \IEEEauthorblockN{Qi Zhang}
% \IEEEauthorblockA{
% % \textit{dept. name of organization (of Aff.)} \\
% \textit{Aarhus University}\\
% % Aarhus, Denmark \\
% qz@ece.au.dk
% }
% % \and
% % \IEEEauthorblockN{4\textsuperscript{th} Given Name Surname}
% % \IEEEauthorblockA{\textit{dept. name of organization (of Aff.)} \\
% % \textit{name of organization (of Aff.)}\\
% % City, Country \\
% % email address or ORCID}
% % \and
% % \IEEEauthorblockN{5\textsuperscript{th} Given Name Surname}
% % \IEEEauthorblockA{\textit{dept. name of organization (of Aff.)} \\
% % \textit{name of organization (of Aff.)}\\
% % City, Country \\
% % email address or ORCID}
% % \and
% % \IEEEauthorblockN{6\textsuperscript{th} Given Name Surname}
% % \IEEEauthorblockA{\textit{dept. name of organization (of Aff.)} \\
% % \textit{name of organization (of Aff.)}\\
% % City, Country \\
% % email address or ORCID}
% }

\maketitle

\begin{abstract}
\pks{The distributed data infrastructure in Internet of Things (IoT) ecosystems requires efficient data-series compression methods, as well as the capability to meet different accuracy demands. However, the compression performance of existing compression methods degrades sharply when calling for ultra-accurate data recovery. In this paper, we introduce \name, a novel highly accurate data compression method that offers a higher compression ratio and lower runtime than prior compressors. \name extracts data \emph{semantics} in the form of linear segments to construct a compact knowledge base, using a dynamic error threshold which can adapt to data characteristics. Then, it captures the remaining data details as \emph{residuals} to support lossy compression at diverse resolutions as well as lossless compression. As \name effectively identifies repeated semantics, its compression ratio increases with data size. Our experimental evaluation demonstrates that \name outperforms state-of-art methods, achieving a twofold to fivefold improvement in compression ratio depending on the dataset.}
\end{abstract}

% The distributed data infrastructure in Internet of Things (IoT) ecosystems requires efficient data-series compression methods, as well as the capability to meet different accuracy demands. However, the compression performance of existing compression methods degrades sharply when calling for ultra-accurate data recovery. In this paper, we introduce SHRINK, a novel highly accurate data compression method that offers a higher compression ratio and lower runtime than prior compressors. SHRINK, extracts data semantics in the form of linear segments to construct a compact knowledge base, using a dynamic error threshold which can adapt to data characteristics. Then, it captures the remaining data details as residuals to support lossy compression at diverse resolutions as well as lossless compression. As SHRINK effectively identifies repeated semantics, its compression ratio increases with data size. Our experimental evaluation demonstrates that SHRINK, outperforms state-of-art methods, achieving a twofold to fivefold improvement in compression ratio depending on the dataset.

\section{Introduction}

Modern Internet of Things (IoT) \emph{edge-based} data infrastructure empowers a distributed paradigm that locates data and computation at the network edge, contrary to traditional \emph{cloud-centric} approaches~\cite{hurst2022glean}. 
% However, to reduce the cost of data storage and improve analytics efficiency~\cite{vestergaard2020randomly}, data stored on resource-constrained edge servers needs to be \emph{compressed}.
Due to limited storage resources at edge servers, data compression is often used to reduce data storage costs~\cite{vestergaard2020randomly}. Whereas lossless compression methods~\cite{blalock2018sprintz, pelkonen2015gorilla, liakos2022chimp} reduce the data volume without incurring information loss, lossy compression methods~\cite{karras2005one, karras2008, chandak2020lfzip, barbarioli2023hierarchical} trade off a small loss of reconstructed data accuracy for higher compression. 
% \gy{Nevertheless, most traditional lossy compression methods fail to efficiently balance compression ratio and data reconstruction accuracy. They degrade sharply when high-accuracy data reconstruction is required, sometimes performing even worse than lossless compression methods.}
Nevertheless, most traditional lossy compression methods fall short of supporting high data reconstructed accuracy, e.g., \(10^\text{-3}\). Their compression performance could degrade dramatically and become even worse than lossless compression when ultra-high accuracy is needed, such as LFZip~\cite{ chandak2020lfzip}, and APCA~\cite{keogh2001locally}.

\begin{figure}[ht]
%\vspace{-3mm}
\centering
\includegraphics[width=\linewidth, height=5cm]{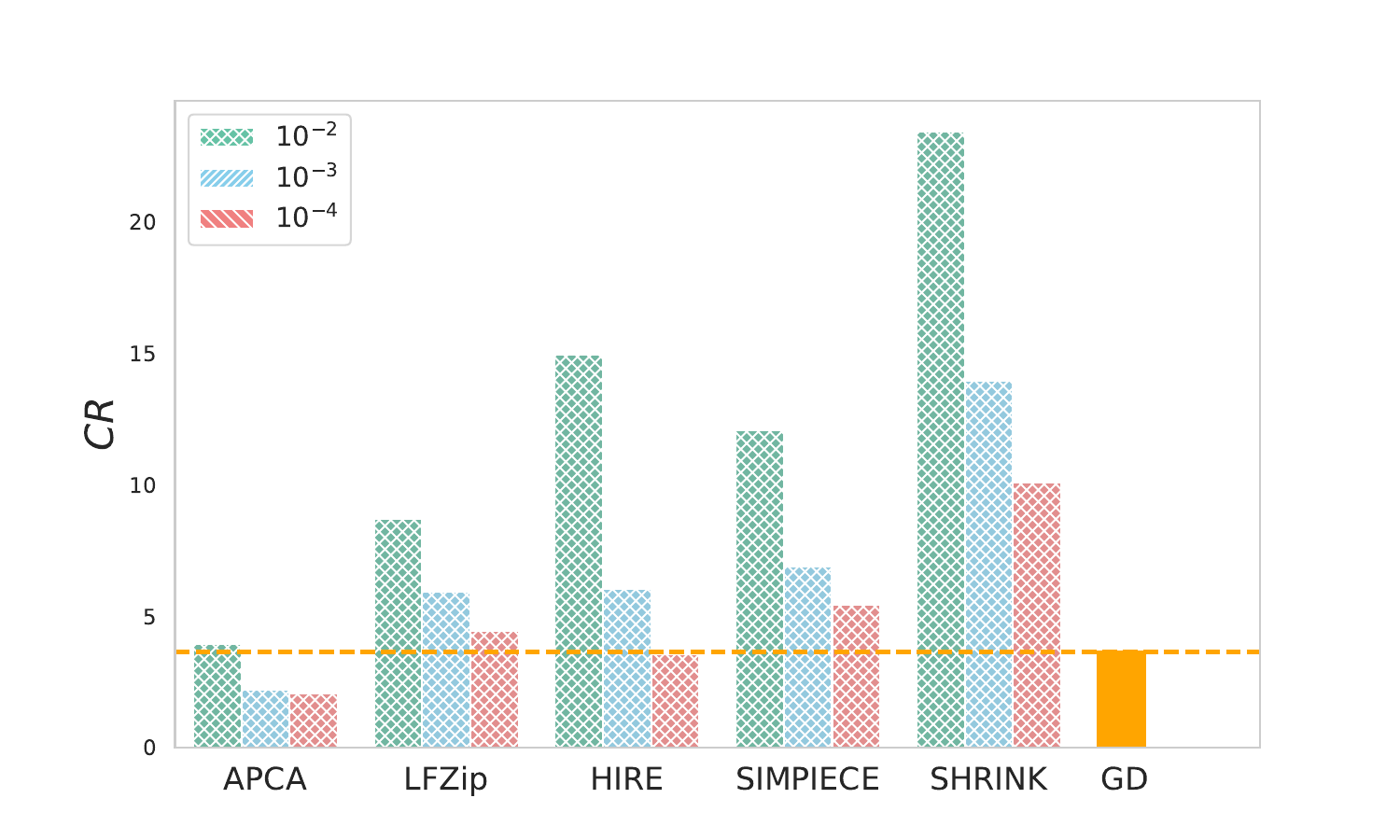}
\vspace{-8mm}
\caption{\pks{Compression ratio of state-of-the-art lossy methods, \name at different~$L_{\infty}$ values and a lossless method (GD).}}\label{fig:Degradation}
\vspace{-3mm}
\end{figure}

% recent advances, problem with lossy schemes
Recent advances in data compression strive to provide sophisticated features in addition to high compression ratio. Lossless compression methods strive to represent the exact data; for instance, Generalised Deduplication (GD)~\cite{vestergaard2020randomly} and \ggd~\cite{hurst2024greedygd} offers random access capability tailored for direct analytics on compressed data with comparable compression ratio as most of general-purpose compressors. Lossy compression methods can provide a lower storage footprint at cost of sacrificing accuracy and are interesting for edge-based data analytics~\cite{paparrizos2021vergedb}. \pks{For instance, \emph{Piecewise Linear Approximation} (PLA) represents data via linear segments, reducing data volume.} A recent lossy compression scheme, \simpiece~\cite{kitsios2023sim}, employs PLA and amalgamates similar line segments by exploiting recurrent data patterns. Nevertheless, when tasked with representing data at high precision, lossy compression schemes falter and yield compression performance worse than lossless compression schemes. \autoref{fig:Degradation} juxtaposes the compression ratios of four lossy methods and one lossless GD method, and our proposal, $\name$. The lossy methods attain high compression and outperform lossless strategies at the modest error threshold of $10^{-2}$, yet their effectiveness degrades rapidly with a strict error tolerance set at~$10^{-4}$.

% what we do
In this paper, we propose \name, a semantic-aware method that achieves ultra-accurate data compression, tailored for IoT edge servers. \name first extracts data semantics in the form of line segments under a base error threshold that adapts to data variability and then merges these semantics into a holistic knowledge base that encodes the underlying data and filters redundancies. Still, these coarse-grained semantics fall short of the applications that require high accuracy. To serve this goal, we augment \name's representation with \emph{residuals}, which drastically reduce bit-level redundancy by virtue of their small variance, contributing to a high compression ratio.

We summarize our main contributions as follows:

% this list need to be progressive and distinguish ideas from results
\begin{enumerate}
\item We reveal that the effectiveness of current \emph{lossy} compression schemes degrades at high accuracy levels.
\item \pks{We propose a two-phase novel compression method, \name that first extracts a knowledge base of semantics capturing enduring data patterns and then augments with residuals expressing transient fluctuations. The core novelty of \name lies in the employment of an \emph{adaptive} error threshold in its semantics extraction phase.}
\item \pks{We show experimentally that \name incurs only a slight increase in the size of the knowledge base as data size grows, meaning an increasing compression ratio for a larger dataset. It achieves up to~5$\times$ higher compression ratios than state-of-the-art methods at a higher throughput, and is especially effective in the case of ultra-accurate compression.}
\end{enumerate}
\begin{table}[!h]
\vspace{-2mm}
\centering
\scriptsize % This reduces the font size
% \footnotesize
\renewcommand{\arraystretch}{1.5}
\caption{Notations}
\vspace{-2mm}
\begin{tabular}{clcl}
\toprule
\multicolumn{1}{c}{\textbf{Symbol}} &\multicolumn{1}{l}{\textbf{Meaning}}& \textbf{Symbol} &\multicolumn{1}{l}{\textbf{Meaning}} \\ \midrule
    $n$              &  Num. of data series                         &$\Delta$         &  Global maximum deviation\\
    $k$              &  Num. of sub-bases                           &$\Delta_i $      & Deviation in interval~$i$    \\
    $S$              & Size of original data						&$L$            &  Default interval length\\
    $S_c$            & Size of compressed data						& $\beta$         & Fluctuation level\\ 
    $S_b$            & Size of base								    & $\lambda $      &  Scaling factor\\ 
    $S_r$            & Size of residuals							& $\origin$       & Origin of shrinking cone\\
    $\epsilon$       &  Error threshold                             & $\Span$         & Span of shrinking cone\\ 
    $\epsilon_b $    &  Base error threshold                        & $\semantic$    &  Semantics of data   \\
    $\epsilon_r$     &  Residual error threshold                    & $B$             & Base of data \\               
    $CR$            &  Compression ratio 							& $R$             & Residuals of data \\
    $X$              &  Original data 							    & $v_i$             & Value of data point at~$i$ \\
    $C_x$            &  Compressed data 							& $\hat{v}_i$       & Approximation of~$v_i$ \\
    \bottomrule
\end{tabular}\label{tab:nomenclature}
\vspace{-2mm}
\end{table}

\section{Problem Formulation}\label{sec:problem}

We now present the fundamental definitions and principles underlying \name. \autoref{tab:nomenclature} lists the main notations.

\subsection{Problem statement}

A data series is a sequence of data points ordered in time order. Typically, given a data series~$X = \left<(t_0, v_0), (t_1, v_1), \ldots, (t_{n-1}, v_{n-1})\right>$ comprising~$n$ data samples, we aim to design a compression method that yields reconstructed data $\hat{X} = \left<(t_0, \hat{v}_0), (t_1, \hat{v}_1), \ldots, (t_{n-1}, \hat{v}_{n-1})\right>$ with a \emph{maximum absolute error} guarantee for each reconstructed data value, i.e., a guarantee by the~$L_{\infty}$ norm, defined as:
\begin{equation}
\epsilon = \lim _{n \rightarrow \infty}\left(\sum_{i=0}^{n-1}\left|\hat{v}_i-v_i\right|^n\right)^{\frac{1}{n}}=\max _i\left|\hat{v}_i-v_i\right| ~.
\end{equation}

We express compressed data in terms of \emph{base}\footnote{In this paper, we use the terms "knowledge base" and "base" interchangeably.} of total size~$S_b$ and \emph{residuals} of total size~$S_r$ and measure compression performance by the \emph{compression ratio}~$CR$, defined as:
\begin{equation}
CR = \frac{S}{S_c} = \frac{S}{S_b+S_r}~,
\end{equation}
\noindent \pk{where~$S$ is the size of the original data and~$S_c$ the size of compressed data, including base and residuals. High values of~$CR$ indicate better performance.}

\begin{figure}[htbp]
\vspace{-3mm}
\centering
\includegraphics[width=\linewidth]{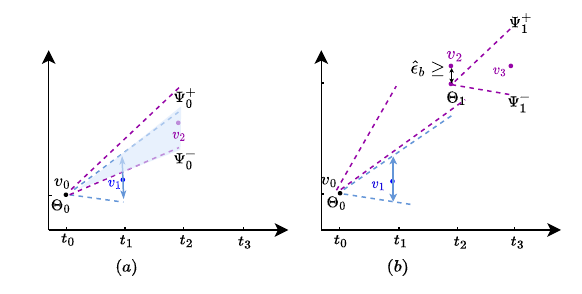}
\vspace{-5mm}
%\caption{With the error threshold, $\name$ clusters data or not. {$\origin_0$} is the origin of both cones. Point 1 is then added. In (a), point 2 (ending key of the gap) is added next, resulting in only one cone. In (b), point 2 is outside the dotted cone and therefore starts a new pattern, yielding different cones with two phases.}
\caption{\pk{Cases of intersecting (a) and disjoint (b) shrinking cones.}}\label{fig:shringkingCone}
\vspace{-3mm}
\end{figure}

\subsection{Semantics of data}

% The aim of extracting \emph{semantics} from data is to identify structures in the data that are meaningful and useful for both (i) compression purposes and (ii) data analytics tasks. While identifying the optimal structures in that regard is an NP-hard problem~\cite{grohe2001evaluation}, particularly for datasets with unidentified characteristics, it suffices to use a sub-optimal structures for compression.

We craft data semantics by \emph{shrinking cones}~\cite{galakatos2019fiting}, elaborated in Section~\ref{sec:Semantics} and~\autoref{fig:shringkingCone}, each of whom clusters points based on their \emph{linear trend}. We thereby represent data by a set~$(B, R, E^*)$, where~$B$ is  \emph{knowledge base} of the dataset, $R$ is the \emph{residuals}, and $E^*$ is a triple of \emph{error thresholds}.

\begin{definition}[Error thresholds]\label{eq:Errorthresholds}
$E^* = \{\epsilon, \epsilon_b, \epsilon_r\}$ is a triple of \emph{error thresholds}, including~$\epsilon_b$ used to extract semantics and build base and~$\epsilon_r$ used to compress residuals. It should be $\epsilon_r \leq \epsilon$, so that reconstructed data are within the error threshold~$\epsilon$.
\end{definition}

\begin{definition}[\emph{Shrinking Cone}]\label{eq:cone}
\pk{A cone is defined by three components: an \emph{origin point}, a \emph{lower slope}, and an \emph{upper slope}, representing a set of viable linear functions with slope between the lower slope and upper slope and starting from the origin point.}
\end{definition}

\begin{definition}[\emph{Base} of data series]\label{eq:base}
The \emph{base} of data ~$B$ sketches the data and by~$k$ disjoint \emph{cones}. Each cone is represented by an \emph{origin}~$\origin$ and a \emph{span}~$\Span$, hence~$B = \left<(\origin_0, \Span_0), (\origin_1, \Span_1), \ldots, (\origin_{k-1}, \Span_{k-1})\right>$, where $\origin_i$ and $\Span_i$ denote the origin and span of sub-base~$B_i$, respectively.
\end{definition}

\begin{definition}[\emph{Origin} of a cone]\label{eq:phase}
\pk{The origin of a cone~$\origin_i$ is the starting point of sub-base~$B_i$, $0 \leq i \leq k-1$, where~$k$ is the number of sub-bases. Cone origins divide a data series into different \emph{phases}. $\origin_i$ is affected by~$\epsilon_b$ and data fluctuation~$\beta$, as we elaborate in Section~\ref{sec:Semantics}.}
\end{definition}

\begin{definition}[\emph{Span} of a cone]\label{eq:trend}
\pk{The \emph{span} of a cone~$\Span_i$ comprises an upper slope~$\Span_i^+$ and a lower slope~$\Span_i^-$. $\Span_i$ represents the slope interval of a linear function that determines the \emph{trend} of the data series at a certain locality and approximates the data points within the cone which share common semantics.}
\end{definition}

\begin{definition}[\emph{Residuals} of data series]\label{eq:residual}
A set of \emph{residuals}~$R$ provides detailed information on a data series obtained by subtraction of the \emph{base}.
\end{definition}

% \autoref{fig:compareForAll} portrays the original data samples, approximation by \emph{base}, \emph{residuals}, and residual kernel density estimation (KDE) plot of an example data series. The \emph{base} provides a sketch of the data series.
As only the essentials, i.e., semantics, are extracted and stored as \emph{base}, the compressed size tends to stay stable regardless of the growth of the total data size. Further, the \emph{residuals} are highly compressible, as they have a small dynamic range and follow a well-behaved distribution. As any data series can be split into knowledge \emph{base} and \emph{residuals}, we build \name based on this property. Details on the computation of base and residuals are provided in the next section.

% \subsection{Application of the method}\label{operation}

% \autoref{fig:edgeframework} depicts the data flow from sensors to edge servers and the cloud and further to data analytics applications. \name compresses data on edge servers, segregating it into base and residuals. The base is stored locally on the edge server, where edge storage space is at a premium, as they encapsulate primary patterns and structures to be used directly for edge-based data analytics (e.g., outlier detection). Residuals, however, are transmitted to the cloud, where storage space is abundant and less expensive, as they represent less critical detailed variations in the data. Ultimately, both base and residuals are available for data analytics, but base can be accessed directly. \name allows downstream applications to perform error-bounded data retrieval, obtaining base from the edge and combining them with residuals from the cloud to reconstruct the full dataset for analysis. As we will see in Section~\ref{edge}, the size of base stays relatively stable, enabling the application of \name in modern IoT edge-based data infrastructures. As the residuals alone do not \gy{provide much useful information}, this data storage framework naturally offers better data protection in the cloud.

% \input{texfigures/edgeframework}
\begin{figure}[!h] 
\vspace{-2mm}
\centering
\includegraphics[width=\linewidth]{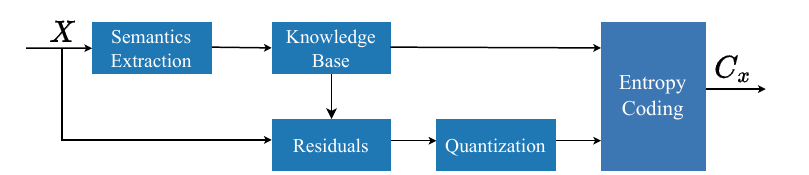}
\vspace{-4mm}
\caption{The workflow of \name.} %The knowledge base with tiny size can be constructed based on $\epsilon_b$ and fluctuation level, and it can be stored either locally or streamed to somewhere else, e.g., a cloud storage server.
\label{fig:Workflow}
\vspace{-2mm}
\end{figure}

\section{Methodology of Shrink}

In this section we describe \name. we give an overview workflow of the \name  in Section~\ref{sec:overview}, present the \emph{adaptive phase division} algorithm that is the foundation of semantics extraction in Section~\ref{sec:Semantics}, detail the process that merges similar semantics into knowledge base in Section~\ref{sec:Base}, describe how we encode residuals to improve compression performance in Section~\ref{sec:residual}.
% , and prove the correctness and complexity of \name in Section~\ref{sec:complexity}.

\vspace{-2mm}
\begin{algorithm}[!h]
\caption{Overall workflow of \name}\label{alg:workflow}
    \begin{algorithmic}[1]
        \State \textbf{Input:}  $X$, $E = $ \{$\epsilon_1$, $\epsilon_2$,...,$\epsilon_n$ \}, $\epsilon_{b}$ \label{lst:line:1}
        \State \textbf{Output:}  $C_x$                                                                        \label{lst:line:2}
        \State $C_x \gets []$                                                                                  \label{lst:line:3}
        \State $ \mathscr{S} \gets \text{SemanticExtraction(X)}$ \Comment{\autoref{sec:Semantics}}              \label{lst:line:4}
        \State $ B_x^{\epsilon_{b}} \gets \text{BaseConstruction($\mathscr{S}$)}$ \Comment{\autoref{sec:Base}}  \label{lst:line:5}
        \State $\hat{B}_x^{\epsilon_{b}} \gets EntropyCoding(B_x^{\epsilon_{b}})$ \Comment{optional}                           \label{lst:line:6}
        \State $C_x.insert(\hat{B}_x^{\epsilon_{b}}) $ 
            \label{lst:line:7}
        \For{ $\epsilon_i$ in $E$}                                                                                                \label{lst:line:8}
                \If{$\epsilon_i \geq \hat{\epsilon}_b$}                                                                         \label{lst:line:9}
                    \State $C_x^i \gets NULL$
                       \label{lst:line:10}  
                \Else $ $ \label{lst:line:11}
                    \State $R_x^{\epsilon_{i}} \gets \text{EncodeResidual}(r, \epsilon_i)$ \Comment{\autoref{sec:residual}}\label{lst:line:12}
                    \State $C_x^i \gets EntropyCoding(R_x^{\epsilon_{i}})$
                    \label{lst:line:13}
                \EndIf                                                                                               \label{lst:line:14}
                \State $C_x.insert(C_x^i) $                                                                               \label{lst:line:15}
                % \State $C_x \gets C_x \cup \{B_x^{\epsilon_{b}} \oplus  R_x^{\epsilon_{i}}\}$                           \label{lst:line:13}
        \EndFor                                                                                                         \label{lst:line:16}
        \State \textbf{return} $C_x$                                                                                     \label{lst:line:17}
    \end{algorithmic}
\end{algorithm}

\vspace{-3mm}

\subsection{Overview}\label{sec:overview}

In \name compressed data is composed of \emph{base} and \emph{residuals} as shown in Equation~\eqref{eq:subencoding}; the~$\oplus$ operation denotes the combination of \emph{base} and \emph{residuals}. \pks{This scheme constructs a single encoding that can be decompressed at various~$L_{\infty}$ error resolutions; this \emph{multiresolution} decompression potential of a single encoding was illustrated in~\cite{barbarioli2023hierarchical}.}

\vspace{-2mm}
\begin{equation}\label{eq:subencoding}
C_x = B_x^{\epsilon_{b}} \oplus R_x^{\epsilon_r} ~.
\end{equation}
\vspace{-2mm}

As \autoref{fig:Workflow} shows, \name (i) extracts semantics adaptively based on data fluctuation, (ii) merges similar semantics to construct base, (iii) encodes residuals to reduce redundancy, (iv) performs entropy coding of quantized residuals and optionally (v) performs entropy coding of base. Algorithm~\ref{alg:workflow} outlines the workflow, which can support multiple applications with diverse error thresholds, $E = \{\epsilon_1, \epsilon_2, \ldots, \epsilon_n \}$ (Lines~\ref{lst:line:1}--\ref{lst:line:2}). \pks{We extract semantics with~$\epsilon_b$ (Line~\ref{lst:line:4}), construct the base (Line~\ref{lst:line:5}) resulting in practical base error threshold~$\hat{\epsilon}_b$, and optionally compress the base using a traditional entropy coding method (Lines~\ref{lst:line:6}--\ref{lst:line:7}). For an~$\epsilon_i$ less demanding than~$\hat{\epsilon}_b$, i.e.,~$\epsilon_i > \hat{\epsilon}_b$, we employ the base ~$B_x^{\epsilon_{b}}$ without residuals} (Lines~\ref{lst:line:8}--\ref{lst:line:10}). Otherwise, we encode residuals and use entropy coding to reduce bits redundancy (Lines~\ref{lst:line:11}--\ref{lst:line:14}). At last, we store each encoded residual to server the apllications (Line~\ref{lst:line:15}). We describe \name considering a univariate data series, yet it also handles multivariate time series by running independently for each dimension.

%Given the diverse error threshold demands for data, {\ourname} only needs to dispatch the \textit{base} instead of the entire dataset repeatedly. This method addresses the issue of redundant data pattern transmission by only sending the minimum necessary semantics. 
%{\ourname} aims to produce a single encoding $C_x$ that can be selectively decomposed into two sub-encodings $B_x^{\epsilon_{b}}$ and $ R_x^{\epsilon_{r}} $ with different error thresholds, as shown in Equation~\ref{eq:subencoding}.

\subsection{Semantics extraction}\label{sec:Semantics}

We consider a data series as a sequence of discrete patterns or semantics, each starting from a data point, or \emph{phase}. The extracted semantics reveal patterns in a dataset~\cite{wang2011finding}. To determine discrete phases, we quantize continuous values to values of fixed precision~\cite{proakis2007digital}. Instead of uniform quantization~\cite{liu2008novel}, we apply \emph{non-uniform quantization}, which rounds each input value differently using an adaptive \emph{quantization step}~\cite{zhou2023adaptive}. Thereby we obtain cone starting points (i.e., origins), possibly shared among cones, which we use to represent cones jointly. The default quantization step for each cone depends on Base error threshold~$\epsilon_b$ and the fluctuation level at the cone's interval. For an interval of length~$L \geq 2$, we define the fluctuation level~$\beta_i = \frac{\Delta_i}{\Delta}$, and set the adaptive quantization step as:

\begin{equation}\label{eq:epsilon}
\hat{\epsilon}_{b,i} = \epsilon_{b} \cdot e^{\frac23 - \beta_i}~,
\end{equation}

\noindent where~$\Delta$ is the global value range of the whole data series, $\Delta_i$ is the local value range in interval~$i$, and~$\epsilon_b$ is the default quantization step. High data fluctuation yields a large~$\beta_i$, hence a small~$\hat{\epsilon}_{b,i}$, hence a more precise quantization to accommodate the greater data variability. Conversely, an interval with low fluctuation leads to a larger~$\hat{\epsilon}_{b,i}$, allowing for a looser quantization threshold since the data does not vary that much. Based on this dynamic quantization step, we first quantize the origin of each cone as follows:

\begin{equation}\label{eq:origin}
\origin_i = \left\lfloor {v_j} \cdot \frac{1} {\hat{\epsilon}_{b,i}} \right\rfloor \cdot {\hat{\epsilon}_{b,i}} ~.
\end{equation}

Algorithm~\ref{alg:phase} outlines this procedure. The default length~$L$ of an interval is set in Lines~\ref{phase:line:4}--\ref{phase:line:5}, controlled by a hyperparameter~$\lambda$ and the default quantization step~$\epsilon_b$. Lines~\ref{phase:line:6}--\ref{phase:line:11} obtain the deviation~$\Delta_i$ and the fluctuation level~$\beta_i$, while Lines~\ref{phase:line:12}--\ref{phase:line:14} derive the actual error threshold~$\hat{\epsilon}_{b,i}$ and set the cone origin based thereupon. Though the default interval length is set to~$L$, the actual length is data-driven.

%\vspace{2mm}
\begin{algorithm}[htbp]																			
    \caption{Phases Division}
    \begin{algorithmic}[1]
    \State \textbf{Input:}  Index of point $j$																				\label{phase:line:1}
    \State \textbf{Output:} Origin of a new cone $\origin$                                                                  \label{phase:line:2}
    \Function{Division}{$j$}                                                                                                \label{phase:line:3}
        \State $L \gets \lambda \cdot n \cdot \epsilon_{b} $                                                                \label{phase:line:4}
        \State $Interval \gets X[j : j+L] $                                                                                 \label{phase:line:5}
        \State $\Delta \gets max - min$                                                                                     \label{phase:line:6}
        \For{ each $v$ in the Interval}                  \label{phase:line:7}
            \State $update(v_{max}, v_{min})$   \label{phase:line:8}
        \EndFor                                                  \label{phase:line:9}
        \State $\Delta_i \gets v_{max} - v_{min}$ \Comment{i is the index of interval}                                      \label{phase:line:10}
        \State $\beta_i \gets \Delta_i/\Delta$                                                                                \label{phase:line:11}
        \State $\hat{\epsilon}_b \gets \epsilon_{b} \cdot e^{(2/3 - \beta_i)}$ 
            \label{phase:line:12}
        \State $ \origin \gets \left\lfloor {v_i}\cdot \frac{1}{\hat{\epsilon}_b} \right\rfloor  \cdot {\hat{\epsilon}_b}$      \label{phase:line:13}
        \State \textbf{return}  $\origin$                                                                                   \label{phase:line:14}
    \EndFunction
    \end{algorithmic}\label{alg:phase}
\end{algorithm}
\vspace{2mm}

Algorithm~\autoref{alg:SemanticsExtraction} illustrates the extraction of semantics. We first set the default bound of cones and quantize the cone origin $\origin_0$ with the dynamic base error threshold~$\hat{\epsilon}_{b}$ (Lines~\ref{se:line:4}--\ref{se:line:7}); If the preceding point's cone does not intersect with the current one, we end the running cone and interval (Lines~\ref{se:line:9}--\ref{se:line:10}) and start a new cone with a new base error threshold from point~$i$ (Lines~\ref{se:line:11}--\ref{se:line:14}). Otherwise, the preceding point's cone intersects the current one, and we update the slopes of span $\Span$ to that intersection (Lines~\ref{se:line:15}--\ref{se:line:17}). When this process terminates, we return the semantics~$\semantic$ (Line ~\ref{se:line:19}). \autoref{fig:shringkingCone} illustrates how we extract data semantics in a cone by a dynamic base error threshold. The cone's upper and lower slopes are set so that any line between them approximates the data points in the interval within~$\hat{\epsilon}_b$. The data interval expands with each newly included data point, leading to further tightening of slope interval~\cite{galakatos2019fiting}, so that lines of slope therein approximate all data points in the data interval within~$\hat{\epsilon}_b$; when the slope interval becomes empty, the expansion terminates. \autoref{fig:shringkingCone}(b) shows an example where there exists no slope interval that can accommodate both the first and second data points observed, hence a new cone starts from point~2. Due to our adaptive base error threshold in Equation~\eqref{eq:epsilon}, when data values in the default interval length vary a little, the cone's span grows and accommodates even more data. Conversely, with high data variability in the default interval length, the cone's span narrows, due to a tighter error margin.

\begin{algorithm}
    \caption{Semantics Extraction}
    \begin{algorithmic}[1]
    \State \textbf{Input:} Data series $X$																				\label{se:line:1}
    \State \textbf{Output:} Semantics  $\semantic$                                                                      \label{se:line:2}
    \Function{SemanticsExtraction}{$X$}                                                                                 \label{se:line:3}
    \State $\semantic \gets []$                                                                                         \label{se:line:4}
    \State $\Span^{+} \gets \infty$                                                                                     \label{se:line:5}
    \State $\Span^{-} \gets -\infty$                                                                                    \label{se:line:6}
    \State $\origin_0 \gets DIVISION({0})$                                                                                \label{se:line:7}
    \For{$(t_i, v_i)$ in $X$}                                                                                                  \label{se:line:8}

        %\If{$v_0<v_i-\hat{\epsilon}_{b}-\Span^{+}(t_i-t_0)$ {or} $v_0>v_i+\hat{\epsilon}_{b}-\Span^{+}(t_i-t_0)$}    
        %\If{$v_0 < v_i - \hat{\epsilon}_{b} - \Span^{+}(t_i - t_0)$ \textbf{or} $v_0 > v_i + \hat{\epsilon}_{b}-\Span^{+}(t_i - t_0)$}
        \If{$\origin_0 < v_i - \hat{\epsilon}_{b} - \Span^{+}\Delta t $ \textbf{or} $\origin_0 > v_i + \hat{\epsilon}_{b} - \Span^{+}\Delta t$}
        \label{se:line:9}
            \State $\semantic.insert([\origin_0, \Span^{-}, \Span^{+},  t_0])$                                       \label{se:line:10}
            \State $\origin_i \gets DIVISION({i})$                                       
            \label{se:line:11}
            \State $(t_0, \origin_0) \gets (t_i, \origin_i)$
            \label{se:line:12}
            \State $\Span^{+} \gets \infty$                                                                             \label{se:line:13}
            \State $\Span^{-} \gets -\infty$                                                                            \label{se:line:14}
        \Else $ $ 
            \State $\Span^{+}  \gets min(\Span^{+}, \frac{v_i+\hat{\epsilon}_{b}-\origin_0}{\Delta t})$  
             \label{se:line:15}
            \State $\Span^{-}  \gets max(\Span^{-},\frac{v_i-\hat{\epsilon}_{b}-\origin_0}{\Delta t})$                                                      \label{se:line:16}
        \EndIf                                                                                                          \label{se:line:17}
    \EndFor                                                                                                             \label{se:line:18}
    \State \textbf{return} $\semantic$                                                                                  \label{se:line:19}
    \EndFunction
    \end{algorithmic}
    \label{alg:SemanticsExtraction}
\end{algorithm}

% \input{texfigures/shringkingCone}

%  The rationale for the adaptive base error threshold is worth mentioning, which is visualized in \autoref{fig:fluctuation}. For volatile data, it is crucial to employ a smaller error bound, which facilitates the retention of data features in the captured semantics and the detection of significant changes. Conversely, for a data stream with low variability, we opt for a larger error bound, as minor fluctuations are often not relevant to the overall analysis and may only represent noise. A larger error bound thus allows for more effective compression while retaining data features.

% \input{texfigures/fluctuation}

\subsection{Base construction}\label{sec:Base}

\begin{figure}[h!]
\vspace{-1mm}
\centering
\includegraphics[width=\linewidth]{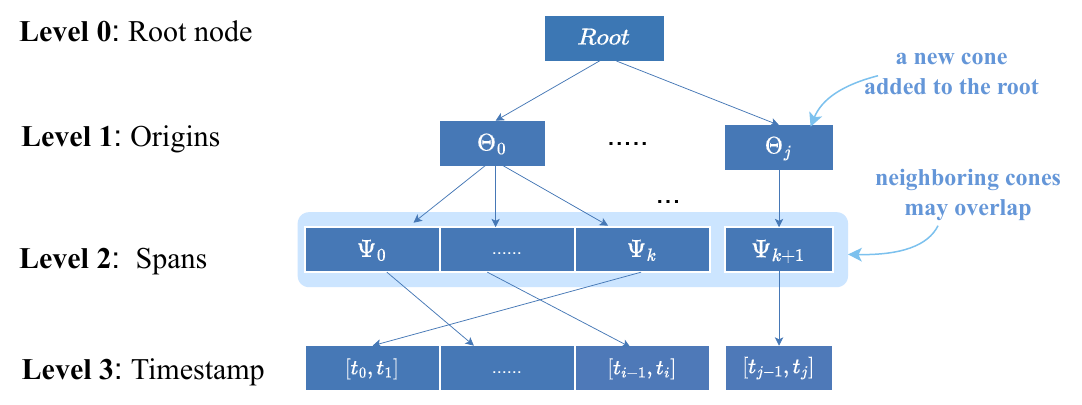}
\vspace{-5mm}
\caption{\pk{Knowledge base construction; overlapping spans sharing a common origin merge to form a base.}}\label{fig:Segments}
\vspace{-1mm}
\end{figure}

To compress data further, we \emph{merge} the extracted semantics based on their similarity.
As we quantize cone origins~$\origin$ to discrete values, it is possible that multiple cones share the same origin.
\autoref{fig:Segments} shows how we order semantics by their origins~$\origin$ and spans~$\Span$ to construct the knowledge base, putting cones in sub-trees. We group cones by their origin and, within each group of the same origin, we order spans in ascending order based on $\Span^{-}$ and serially scan the sorted list to greedily detect contiguous groups of cones with intersecting spans, which we merge and represent compactly; the ensuing segmentation minimizes groups~\cite{karras07, kitsios2023sim}. We thus build a knowledge base~$B = \left\{B_1, B_2, \cdots, B_k\right\}$ by the similarity of~$\origin$ and~$\Span$.

Algorithm~\autoref{alg:BaseConstruction} shows the workflow of semantics merging and knowledge base construction. It organizes cones in a priority queue, based on their origin and lower slopes, in ascending value of  ${\origin}$ and $\Span^-$ (Lines~\ref{BC:line:1}--\ref{BC:line:7}). Cones sharing the same origin are placed in the same sub-tree, as Level 1 of \autoref{fig:Segments} shows. In each sub-tree, we first initialize the default \emph{subbase}  (Line~\ref{BC:line:8}--\ref{BC:line:9}).Then, we iterate over each span ${\Span_j}$ with the same origin ${\origin_i}$ (Line~\ref{BC:line:10}). If the current span overlaps with the span of \emph{subbase}, we update \emph{subbase} with the intersection of the two spans and add the timestamp accordingly (Line~\ref{BC:line:11}--\ref{BC:line:14}). Otherwise, we add the \emph{subbase} into the knowledge base ~$B_x^{\epsilon_{b}}$ and update the \emph{subbase} with the current cone (Line~\ref{BC:line:15}--\ref{BC:line:17}). When finishing the traverse and merge of all the origins, we return the knowledge base ~$B_x^{\epsilon_{b}}$ (Line ~\ref{BC:line:21}). For example, in ~\autoref{fig:Segments}, we merge $\Span_0 = [\Span^-_0, \Span^+_0]$ with its neighbor~$\Span_1 = [\Span^-_1, \Span^+_1]$, into one cone if~$\Span^-_1 \geq \Span^+_0$, and continue with neighboring cones. This merging process ensures an optimal result with a perfect elimination scheme~\cite{kitsios2023sim, gupta1982efficient}. When this merging process terminates, the knowledge base ~$B_x^{\epsilon_{b}}$ is constructed.

\vspace{-1mm}
% \begin{algorithm}
% \caption{Base Construction}
% \begin{algorithmic}[1]
% \State \textbf{Input:} $\semantic$ 
% \State \textbf{Output:} $B_x^{\epsilon_{b}}$
% \Function{BaseConstruction}{$Semantics$}
%     \State $Root \gets PriorityQueue()$ 
%      \For{$c$ in $\semantic$ }
%         \State $Root.append($c$)$
%       \EndFor
%       \State $B_x^{\epsilon_{b}} \gets Merge(Root)$
%     \State \textbf{return} $B_x^{\epsilon_{b}}$
% \EndFunction
% \end{algorithmic}
% \label{alg:BaseConstruction}
% \end{algorithm}

\begin{algorithm}
\caption{Base Construction}
\begin{algorithmic}[1]
\State \textbf{Input:} $\semantic$ 						 \label{BC:line:1}
\State \textbf{Output:} $B_x^{\epsilon_{b}}$			 \label{BC:line:2}
\Function{BaseConstruction}{$\semantic$}                 \label{BC:line:3}
    \State $Root \gets PriorityQueue()$                   \label{BC:line:4}
     \For{$c$ in $\semantic$ }                           \label{BC:line:5}
        \State $Root.insert($c$)$    \Comment{order by $\origin$ and $\Span^{-}$}                                      \label{BC:line:6}
      \EndFor                                            \label{BC:line:7}
      \For{$\origin_i$ in $Root$ }                           \label{BC:line:8}
            \State $subbase \gets [\Span^{-}=-\infty,\Span^{+}=\infty,t=NULL]$ \label{BC:line:9}
            \For{$\Span_j$ in $\origin_i$ } \label{BC:line:10}
                    \If{$\Span_j^{-}\leq subbase.\Span^{+}$ \textbf{and} $\Span_j^{+}\geq subbase.\Span^{-}$}  \label{BC:line:11}
                            \State $subbase.\Span^{-} \gets max(\Span_j^{-}, subbase.\Span^{-})$ \label{BC:line:12}
                            \State $subbase.\Span^{+} \gets min(\Span_j^{+}, subbase.\Span^{+})$ \label{BC:line:13}
                            \State $subbase.t.append(t_j) $ \label{BC:line:14}
                        \Else $ $ \label{BC:line:15}
                            \State $B_x^{\epsilon_{b}}.insert([\origin_i,subbase])$  \label{BC:line:16}
                            \State $subbase \gets [\Span^{-}=\Span_j^{-},\Span^{+}=\Span_j^{+},t=t_j]$ \label{BC:line:17}
                        \EndIf\label{BC:line:18}
            \EndFor \label{BC:line:19}
      \EndFor \label{BC:line:20}                                           
     \State \textbf{return} $B_x^{\epsilon_{b}}$          \label{BC:line:21}
\EndFunction
\end{algorithmic}
\label{alg:BaseConstruction}
\end{algorithm}
\vspace{1mm}

\subsection{Residuals encoding}\label{sec:residual}

\pks{While the derived knowledge base preserves critical data features within error~$\hat{\epsilon}_b$ and eliminates redundancies, it does not suffice to yield the high reconstruction accuracy required by some applications~\cite{aguerrebere2023similarity}. To enhance reconstruction accuracy, we use \emph{residuals}.}

\begin{figure}[h!]
\vspace{-2mm}
\centering
\includegraphics[width=\linewidth]{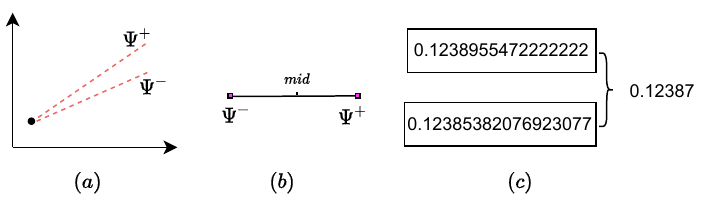}
\vspace{-8mm}
\caption{\pk{(a) slopes; (b) candidate middle; (c) truncated average.}}\label{fig:averageSlope}
\vspace{-3mm}
\end{figure}

For a cone represented by~$(\origin_i, \Span_i^-, \Span_i^+)$, any line of slope between~$\Span^-$ and~$\Span^+$ suffices to represent all underlying data. Conventional piece-wise linear approximation uses the line of slope~ $\frac{\Span^{-} + \Span^{+}}2$. However, the exact average conveys unnecessarily high precision. As \autoref{fig:averageSlope} shows, a cone with~$\Span^{-} = 0.12385382076923077$ and~$\Span^{+} = 0.1238955472222222$ yields average slope~$0.12387468399572649$. We opt to use 5 digits of precision, representing the slope as~$0.12387$ without significant loss of accuracy.

Algorithm~\autoref{alg:CandidateLine} presents our algorithm for slope selection. In case~$\Span^{-}$ and~$\Span^{+}$ have different integer parts, implying that the cone has a quite big span, we retain an average slope with a single decimal digit (Lines~\ref{CL:line:4}--\ref{CL:line:8}). Otherwise, we retain the common decimal values of~$\Span^{-}$ and~$\Span^{+}$, prefixed with the common integer part and suffixed with the average of their first divergent digits (Lines~\ref{CL:line:9}--\ref{CL:line:12}). The candidate line of the current sub-base is equipped with the desired slope (Line \ref{CL:line:13}). After we construct the knowledge base and choose the desired line, we calculate \emph{residuals}, i.e., the difference between each original value and its produced approximation. Since residuals are expressed by cones, they range within $[-\hat{\epsilon}_b, \hat{\epsilon}_b]$. We quantize residuals using~$\epsilon_r$ as quantization step, with~$\epsilon_r \leq \epsilon$.

\vspace{2mm}
% \begin{algorithm}
% \caption{Candidate Line Selection}
% \begin{algorithmic}[1]
% \State \textbf{Input:}  $\Span^{-}$, $\Span^{+}$
% \State \textbf{Output:} Slope $\mu$ of candidate line
% \Function{OptimizedSlope}{$\Span^{-}$, $\Span^{+}$}
%     \State $lead_1 \gets \Call{ExtractIntegerPart}{\Span^{-}}$
%     \State $lead_2 \gets \Call{ExtractIntegerPart}{\Span^{+}}$
%     \If{$lead_1 \neq lead_2$}
%         \State \Return $\Call{Round}{(\Span^{-} + \Span^{+}) / 2, 1}$
%     \EndIf
%     \State $tail_1 \gets \Call{ExtractDecimalPart}{\Span^{-}}$
%     \State $tail_2 \gets \Call{ExtractDecimalPart}{\Span^{+}}$
%     \State $tail \gets \Call{LongestCommonPrefix}{tail_1, tail_2}$
%     \State $\mu \gets \Call{Concatenate}{lead_1, '.', tail}$
%     \State \textbf{return} $\mu$
% \EndFunction
% \end{algorithmic}
% \label{alg:CandidateLine}
% \end{algorithm}

\begin{algorithm}
\caption{Candidate Line Selection}
\begin{algorithmic}[1]
\State \textbf{Input:}  $\Span^{-}$, $\Span^{+}$						\label{CL:line:1}
\State \textbf{Output:} Slope $\mu$ of candidate line					 \label{CL:line:2}
\Function{OptimizedSlope}{$\Span^{-}$, $\Span^{+}$}                      \label{CL:line:3}
    \State $lead_1 \gets \Call{ExtractIntegerPart}{\Span^{-}}$           \label{CL:line:4}
    \State $lead_2 \gets \Call{ExtractIntegerPart}{\Span^{+}}$           \label{CL:line:5}
    \If{$lead_1 \neq lead_2$}                                            \label{CL:line:6}
        \State \Return $\Call{Round}{(\Span^{-} + \Span^{+}) / 2, 1}$    \label{CL:line:7}
    \EndIf                                                               \label{CL:line:8}
    \State $tail_1 \gets \Call{ExtractDecimalPart}{\Span^{-}}$           \label{CL:line:9}
    \State $tail_2 \gets \Call{ExtractDecimalPart}{\Span^{+}}$           \label{CL:line:10}
    \State $tail \gets \Call{LongestCommonPrefix}{tail_1, tail_2}$       \label{CL:line:11}
    \State $\mu \gets \Call{Concatenate}{lead_1, '.', tail}$             \label{CL:line:12}
    \State \textbf{return} $\mu$                                         \label{CL:line:13}
\EndFunction
\end{algorithmic}
\label{alg:CandidateLine}
\end{algorithm}

\vspace{-2mm}

% The residual differences between approximate and original values have small amplitudes yet diverse value frequencies~\cite{wang2022frequency}. While residuals do not significantly contribute to the knowledge base, they requires large number of bits to be stored. They follow, however, a distribution with a mean close to zero, as shown in~\autoref{fig:compareForAll}(d). We leverage these characteristics to quantize each residual~$r_i$ by a \emph{residual quantization step}~$\epsilon_r < \epsilon$ and round floating-point values down to the nearest integer:

The residual differences between approximate and original values have small amplitudes yet diverse value frequencies~\cite{wang2022frequency}. While residuals do not significantly contribute to the knowledge base, they requires large number of bits to be stored. They follow, however, a distribution with a mean close to zero. We leverage these characteristics to quantize each residual~$r_i$ by a \emph{residual quantization step}~$\epsilon_r < \epsilon$ and round floating-point values down to the nearest integer:

\begin{equation}
\operatorname{Q}\left(r_i\right) = \left\lfloor \frac{r_i - r^-}{\epsilon_r} \right\rfloor
\end{equation}

In effect, we obtain rounded integer values in the range~$\left[ 0, \left\lfloor\frac{r^+ - r^-}{\epsilon_r}\right\rfloor \right]$, where~$r^{-}$ and~$r^{+}$ are the minimum and maximum residual values, respectively. We further improve the compression ratio using \emph{entropy coding}. We use Turbo Range Coder (TRC), an arithmetic encoder built on top of the Burrows–Wheeler transform~\cite{burrows1994block} to reorganize blocks of values into sequences of identical digits. We can combine the residuals with the base in data decompression to achieve highly accurate data recovery. Algorithm~\autoref{alg:ResidualEncoding} details the residual encoding process. We first initialize an empty list to store the residuals (Line~\ref{RE:line:4}). Then, we iterate over each sub-base $b_i$ (Line~\ref{RE:line:5}). The slope of the candidate line in $b_i$ is obtained in (Line~\ref{RE:line:6}). With the slope, we compute the residuals related to $b_i$ accordingly and put it into $R$ (Lines~\ref{RE:line:7}--\ref{RE:line:8}). Based on $\epsilon_r$, we quantize the residuals to reduce redundancy and return the quantized one (Lines~\ref{RE:line:10}--\ref{RE:line:11}).

%In theory, it would result in a storage size of $\log _2{\left\lceil\frac{1}{\hat{\epsilon}_{r}}\right\rceil}$ every integer with simple bit-level encoding. Motivated by Sprintz ~\cite{blalock2018sprintz}, we could use vectorization to store more residual rather than the value one by one, because we know much of the residuals are similar.

\vspace{-2mm}
% \begin{algorithm}
% \caption{Residuals Encoding}
% \label{alg:residualencode}
% \begin{algorithmic}[1]
% \State \textbf{Input:}  $B_x^{\epsilon_{b}}$
% \State \textbf{Output:} $r_x^{\epsilon_{r}}$
% \Function{EncodeResiduals}{$B_x^{\epsilon_{b}}$}

%         \For{each $b$ in $B_x^{\epsilon_{b}}$} 
%             \State $\Span_b \gets OptimizedSlope(\Span^{-}_b, \Span^{+}_b)$
%             % \State $B \gets Decompose(B_x^{\epsilon_{b}})$
%             \State $r \gets X - B_x^{\epsilon_{b}}$
%             \State $\hat{r} \gets Quantiz(r)$
%             \State $R \gets R \cup \{\hat{r} \}$
%         \EndFor
%         \State $r_x^{\epsilon_{r}}  \gets EntropyCoding(R)$
        
%     \State \textbf{return} $r_x^{\epsilon_{r}}$
% \EndFunction
% \end{algorithmic}
% \label{alg:ResidualEncoding}
% \end{algorithm}

\begin{algorithm}
\caption{Residuals Encoding}
\label{alg:residualencode}
\begin{algorithmic}[1]
\State \textbf{Input:}  $B_x^{\epsilon_{b}}, \epsilon_r	$							\label{RE:line:1}
\State \textbf{Output:} $R_x^{\epsilon_{r}}$								 \label{RE:line:2}
\Function{EncodeResiduals}{$B_x^{\epsilon_{b}}, \epsilon_r$}                             \label{RE:line:3}
		\State $R \gets [ ]$																	 \label{RE:line:4}
        \For{each $b_i$ in $B_x^{\epsilon_{b}}$}                               \label{RE:line:5}
            \State $\Span_{b_{i}} \gets OptimizedSlope(\Span^{-}_{b_{i}}, \Span^{+}_{b_{i}})$  \label{RE:line:6}
            % \State $B \gets Decompose(B_x^{\epsilon_{b}})$                 \label{RE:line:7}
            \State $r_{b_{i}} \gets X_{b_{i}} - sketch(b_i, \Span_{b_{i}})$                          \label{RE:line:7}
            \State $R.append(r_{b_{i}})$                          \label{RE:line:8}
        \EndFor                                             \label{RE:line:9}
            \State $R_x^{\epsilon_{r}} \gets Quantize(R, \epsilon_{r})$                                \label{RE:line:10}
    \State \textbf{return} $R_x^{\epsilon_{r}}$                              \label{RE:line:11}
\EndFunction
\end{algorithmic}
\label{alg:ResidualEncoding}
\end{algorithm}
\vspace{2mm}
\section{Experimental Evaluation}

We evaluate the performance of \name on a common computing system equipped with an Intel i7-10510U processor, 16GB of RAM, and a 256GB solid-state drive. The algorithm was implemented in Python version~3.9.16. We use the Turbo Range Coder to encode residuals into bytes~\cite{turboRangeCoder}. We assess performance on a suite of five data series from the UCR time series data repository~\cite{ucr}, including FaceFour, MoteStrain, Lightning, Cricket, and Wafer, as well as four more datasets, each exemplifying unique patterns of variability and trend, sourced from the National Ecological Observatory Network (Wind Speed, Wind Direction, and Pressure) and human electrocardiogram data (ECG) data~\cite{ecg}. \autoref{tab:dataset} provides the details of these datasets.

% \begin{table}[h!]
%      \footnotesize
%      \vspace{-1em}
%      \scriptsize % This reduces the font size
%      \renewcommand{\arraystretch}{1.5}
%     \caption{Datasets\tablefootnote{\pk{Decimal means max decimal places; $\max$ and $\min$ rounded to one decimal place.}} used for evaluation}
%     \vspace{-0.5em}
%     \begin{tabular}{lcrrrr}
%     \toprule
%     \textbf{Dataset} & \textbf{Decimal} & \textbf{Max} & \textbf{Min} & \textbf{Num. rows} & \textbf{Size (MB)} \\ \midrule
%     FaceFour \cite{ucr}         & 8         & 5.9        & -4.6       & 39\,200     & 0.67              \\
%     MoteStrain \cite{ucr}       & 8         & 8.5        & -8.5       & 106\,848    & 1.85              \\
%     Lightning \cite{ucr}        & 8         & 23.1       & -1.6       & 122\,694    & 2.19              \\
%     ECG \cite{ecg}              & 11        & 7.4        & -7.0        & 699\,720   & 12.02             \\
%     Cricket \cite{ucr}          & 8         & 12.7       & -10.1      & 702\,000    & 12.78             \\
%     Wind Dir. \cite{neon}       & 2         & 360.0      & 0.0        & 4\,119\,081   & 16.35             \\
%     Wafer \cite{ucr}            & 7         & 12.1       & -3.0       & 1\,088\,928   & 19.64             \\
%     Wind Speed \cite{neon}      & 2         & 20.4       & 0.0        & 1\,169\,510   & 53.23             \\
%     Pressure \cite{pressure}    & 5         & 104.1      & 90.9       & 12\,098\,677  & 214.79            \\  \bottomrule
%     \end{tabular}
%     \vspace{-0.3em}
%     \label{tab:dataset}
% \end{table}

\begin{table}[h!]
     \footnotesize
     \vspace{-1em}
     \scriptsize % This reduces the font size
     \renewcommand{\arraystretch}{1.5}
    \caption{Datasets\tablefootnote{\pk{Decimal means max decimal places; $\max$ and $\min$ rounded to one decimal place.}} used for evaluation}
    \vspace{-0.5em}
    \begin{tabular}{lcrrrr}
    \toprule
    \textbf{Dataset} & \textbf{Decimal} & \textbf{Max} & \textbf{Min} & \textbf{Num. rows} & \textbf{Size (MB)} \\ \midrule
    FaceFour          & 8         & 5.9        & -4.6       & 39\,200     & 0.67              \\
    MoteStrain        & 8         & 8.5        & -8.5       & 106\,848    & 1.85              \\
    Lightning         & 8         & 23.1       & -1.6       & 122\,694    & 2.19              \\
    ECG               & 11        & 7.4        & -7.0        & 699\,720   & 12.02             \\
    Cricket           & 8         & 12.7       & -10.1      & 702\,000    & 12.78             \\
    Wind Dir.        & 2         & 360.0      & 0.0        & 1\,169\,510   & 16.35             \\
    Wafer             & 7         & 12.1       & -3.0       & 1\,088\,928   & 19.64             \\
    Wind Speed       & 2         & 20.4       & 0.0        & 4\,119\,081  & 53.23             \\
    Pressure     & 5         & 104.1      & 90.9       & 12\,098\,677  & 214.79            \\  \bottomrule
    \end{tabular}
    \vspace{-0.3em}
    \label{tab:dataset}
\end{table}

Extensive experiments were performed to compare \name with PLA method \simpiece~\cite{kitsios2023sim}, which demonstrated better performance than other counterparts, such as Mixed-PLA~\cite{luo2015piecewise}, Swing and Slide~\cite{elmeleegy2009online}. APCA~\cite{keogh2001locally} was also included because it adopts a different piecewise constant segment method. The popular lossless compression methods(i.e., Bzip2~\cite{bzip2site},  GZip~\cite{gzipsite}, TRC~\cite{turboRangeCoder}, Gorilla~\cite{pelkonen2015gorilla}, GD~\cite{hurst2024greedygd}) and general-purpose lossy ones (i.e., HIRE~\cite{barbarioli2023hierarchical}, LFZip~\cite{chandak2020lfzip}) were also included.

\subsection{Results on compression ratio}\label{sec:cr}

\pks{In this section, we evaluate the compression ratio of \name against (i) piecewise-segment-based lossy compression methods, (ii) general-purpose lossy compression methods, and (iii) state-of-the-art lossless compression methods.}

\subsubsection{Piecewise-segment lossy compression}\label{sec:picewise}

We first present a detailed comparative analysis of compression ratios against two representative \emph{lossy} piecewise segment compression methods, \simpiece and APCA, under nine error resolution levels that are inside the scope of real world usage, \{0.01, 0.0075, 0.005, 0.0025, 0.001, 0.00075, 0.0005, 0.00025, 0.0001\}. For the datasets of Windspeed and Wind Direction, the error resolution levels are set to \{0.01, 0.0075, 0.005, 0.0025, 0.001\}, because these datasets only have two decimals for each data point. We choose these error thresholds, given that industrial stakeholders are interested in compression at high rather than low precision, even though most lossy compression methods in the literature offer compression at low precision. \name addresses this gap. \pks{We extract semantics setting the error threshold~$\epsilon_b$ at~5\% of the dataset range.}

\begin{figure}[!ht]
\vspace{-2mm}
\centering
\includegraphics[width=\linewidth]{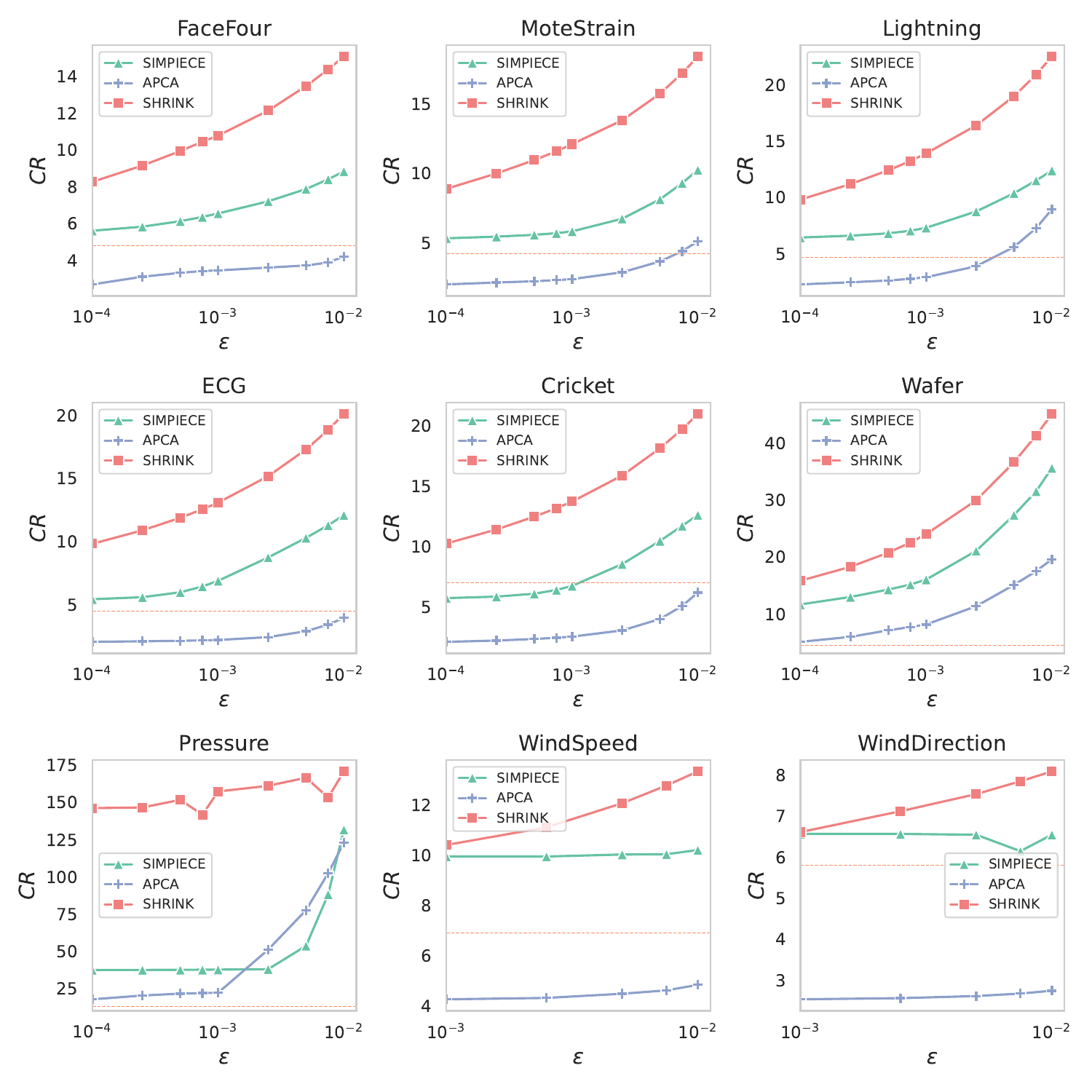}
\vspace{-7mm}
\caption{\pks{Comparison to lossy piecewise segment methods; the dashed line indicates the compression ratio of lossless \name.}}\label{fig:crlossy}
\vspace{-2mm}
\end{figure}

\autoref{fig:crlossy} presents the detailed compression ratios of \name, \simpiece, and APCA, while the dashed line indicates the compression ratio of lossless \name. \name surpasses \simpiece and APCA on all the datasets. Interestingly, lossless \name achieves a higher compression than lossy \simpiece and APCA on some datasets, e.g., Cricket. Further, \name achieves a much higher compression ratio than \simpiece and APCA on larger data sets, e.g.,~150$\times$ to~170$\times$ compression on the Pressure dataset. Besides, \simpiece outperforms APCA on all datasets except the Pressure dataset; that data set presents frequent identical consecutive values, in which APCA, as a method tailored for piecewise constant approximation method, gains more.
% \pks{Notably, even while \name outperforms \simpiece for the WindSpeed and WindDirection dataset in nearly all error thresholds, it performs similarly to \simpiece at the strict error threshold of~$\epsilon = 10^{-3}$. We attribute this behavior to the characteristics of wind data, which feature sharp discontinuities. Such discontinuities pose a challenge to methods using linear approximation. Still, even though \name cannot capture sufficient semantics in such a data set with a small~$\epsilon_b$, it allows adjusting the base error threshold to improve performance, as we outline in Section~\ref{Sec:granularity}.}

\subsubsection{General-purpose lossy compression}

\begin{figure}[!ht]
\centering
\vspace{-2mm}
\includegraphics[width=\linewidth]{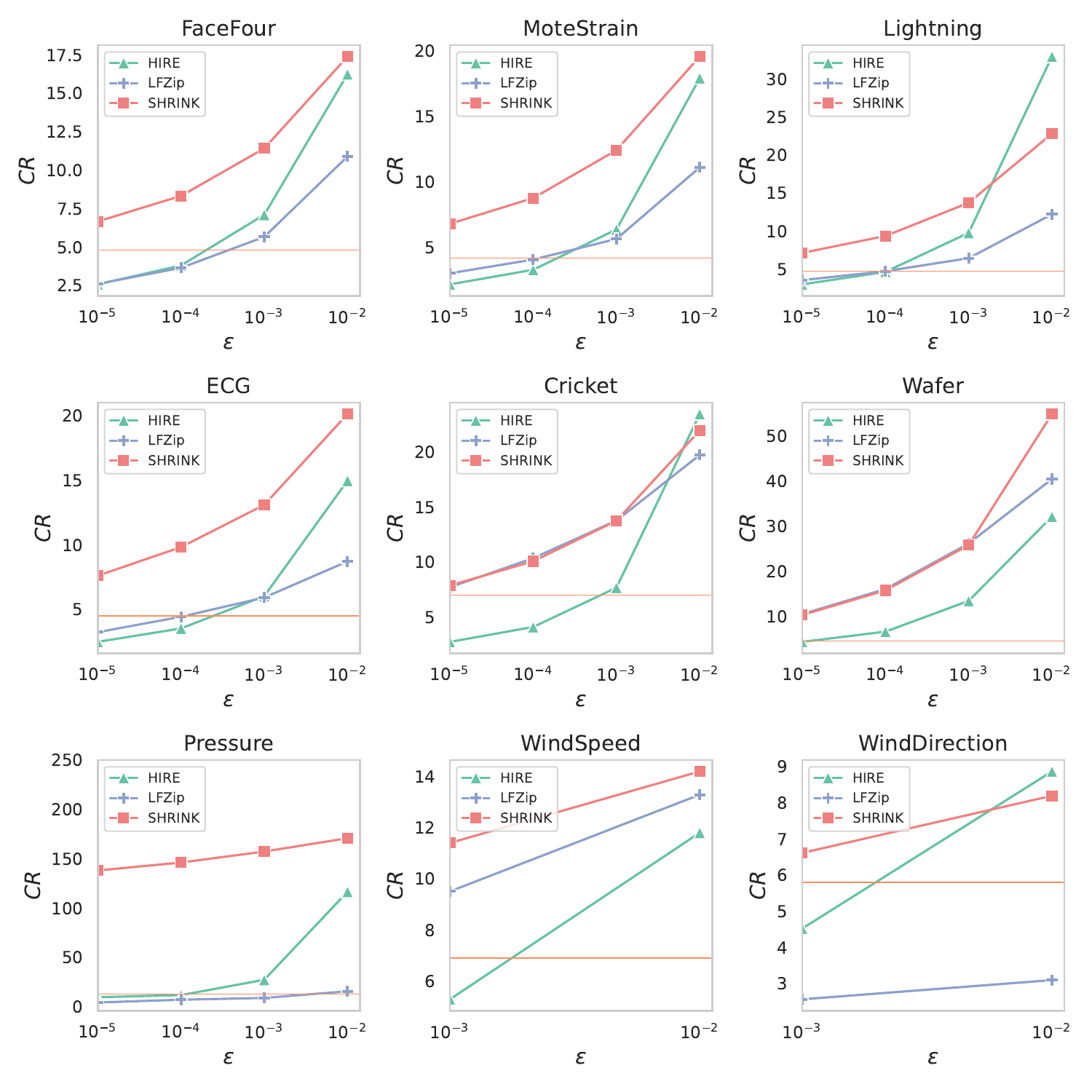}
\vspace{-7mm}
\caption{\pks{Comparison vs. general-purpose lossy compression.}}
\label{fig:multiresolution}
\vspace{-2mm}
\end{figure}

Next, we evaluate the performance of \name against two general-purpose lossy compression methods, HIRE and LFZip, using error resolution levels ranging from~$10^{-2}$ to~$10^{-5}$ on a logarithmic scale. \pks{This broad range ensures a comprehensive evaluation, reflecting the stringent precision requirements and diverse application scenarios of general-purpose methods.} For WindSpeed and WindDirection datasets, the error resolution is set from~$10^{-2}$ to $10^{-3}$ due to limited decimal places in the datasets. We set~$\epsilon_b$ as~15\% of the data range since the compression performance is the main goal for general purpose compression. \autoref{fig:multiresolution} presents the experimental results, in which the dashed line indicates the lossless compression ratio achieved by \name. Our results demonstrate that \name consistently outperforms HIRE and LFZip across nearly all datasets and error thresholds, achieving higher compression ratios while maintaining data accuracy, particularly at stringent error thresholds, i.e., ~$\epsilon \le 10^{-3}$.

\pks{We emphasis that the setting of~$\epsilon_b$, and also our the selection of error thresholds, differs from that in Section~\ref{sec:picewise} due to the distinct objectives of the compared methods. Piecewise-segment compression methods (e.g., \simpiece and APCA) aim to capture and mine meaningful structures in the data, where error resolutions play a cardinal role. With a stringent error resolution, e.g.,~$\epsilon \leq 10^{-4}$, those methods' compression performance deteriorates, as they fail to capture meaningful structures in the data. Contrariwise, \name performs well even at stringent error resolutions by virtue of its \emph{adaptable} error threshold in its semantics extraction phase; this adaptable error threshold allows \name to capture meaningful structures at a laxer error resolution before adding residual to attain higher accuracy. Our results show that this error-adaptation strategy attains high compression even at stringent error resolutions.}

%\gy{The selection of error thresholds differs from those in Section~\ref{sec:picewise} due to the distinct objectives and natures of the compared methods. Lossy piecewise segment compression methods (e.g., \simpiece and APCA), in particular, piecewise linear approximation, aim to capture meaningful structure or mine datasets based on a representation of data~\cite{keogh2001online}. Thus, different error thresholds are significant as traditional PLA methods do not adjust their error thresholds adaptively to capture more meaningful structures or semantics of data. Therefore, we use a more detailed error resolution rather than a broader one for PLA methods. On the other hand, general-purpose compression methods are designed to perform efficiently across a broader spectrum of error tolerances to achieve higher compression ratios. The broader error resolution range selected for this sub-section ensures that the evaluation reflects typical use cases for these versatile methods while maintaining rigorous precision. Thus, we do not need to use as fine-grained a comparison for general-purpose methods. However, SHRINK strikes a good balance between them, as it not only effectively captures meaningful semantics but also adjusts errors in a broader range to improve its performance.}

\subsubsection{Lossless compression}

% Finally, to evaluate \name on \emph{lossless} compression, \pks{we compare to three general-purpose compressors: BZip2, GZip, and TRC, and two specific-purpose compressors: GD and Gorilla, on the same nine datasets.} 

\begin{figure}[h!]
\vspace{-3mm}
\centering
\includegraphics[width=0.9\linewidth]{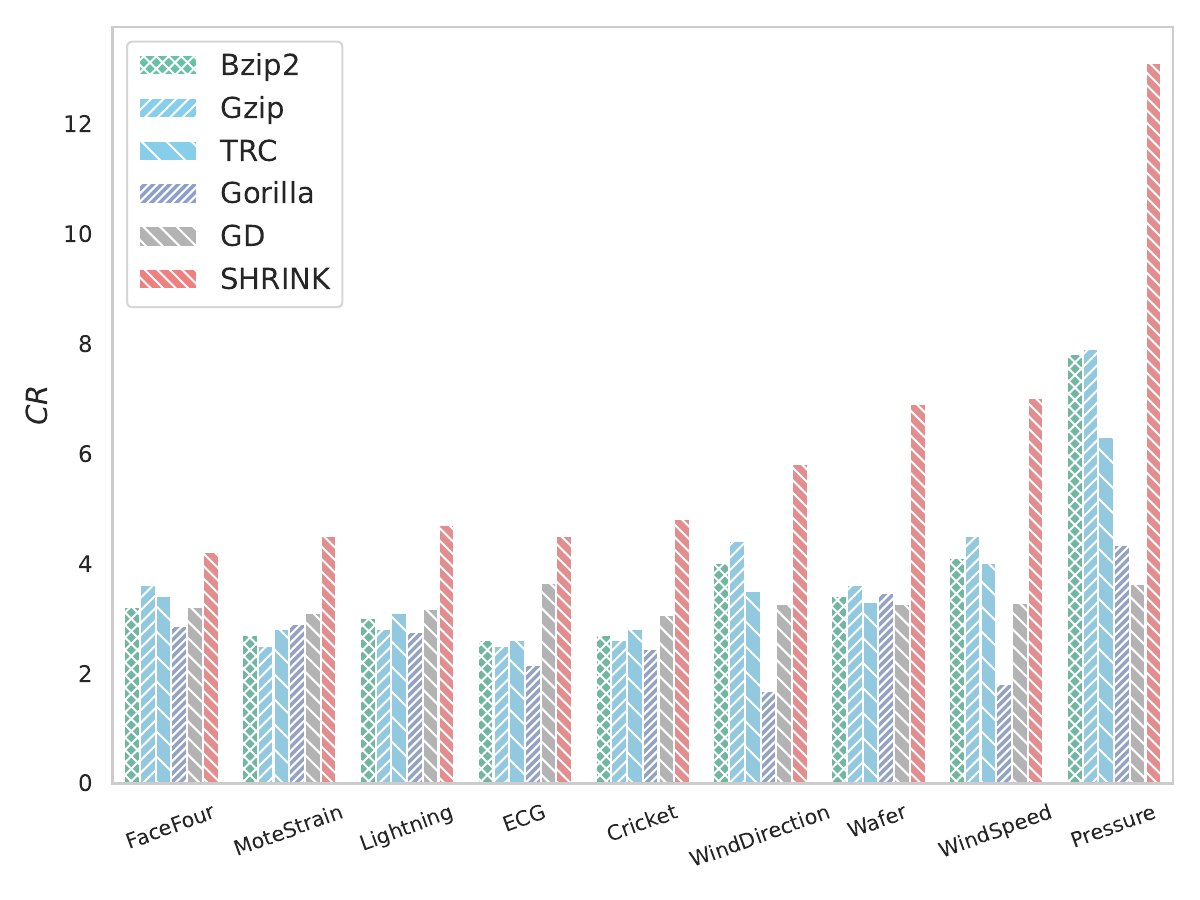}
\vspace{-6mm}
\caption{\pks{Lossless compression ratio on 9 datasets.}}\label{fig:crlossless}
\vspace{-4mm}
\end{figure}

\autoref{fig:crlossless} depicts evaluation on \emph{lossless} compression. \name outperforms all competitors here too, \pks{with an up to twofold improvement}. Notably, these conventional techniques merely perform \emph{bit-level} compression without considering the data semantics. Contrarily, \name leverages the intrinsic features and correlations within the data, thereby furnishing a more effective compression. Remarkably, \name achieves a compression of more than~12$\times$ on the Pressure dataset. The particularly high compression ratio on a dataset with complex data reconfirms that \name's performance scales with the complexity and size of the data. This observation entails that \name achieves more effective data reduction on larger datasets, a significant advantage in applications that require storing very large data series. \pks{Besides, general-purpose compressors perform generally better than the two specific-purpose ones; the special attention of the latter to specific purposes, such as random access for GD and streaming compression for Gorilla, compromises compression ratio slightly for some datasets.}

% default~$\epsilon_b = 5\% * range$. With the Wind Direction data, we relax~$\epsilon_b$ to~10\% of the value range, as this data series has a larger value range than others and behaves differently, with sharp increases and decreases, rendering it unnecessary to set a small~$\epsilon_b$.

% \autoref{fig:CRLossyaverage} shows our results on the average compression ratio over all error thresholds. {\name} achieves from~2.5$\times$ to~1.5$\times$ higher compression ratio than APCA and \simpiece.
% \input{texfigures/CRLossyaverage}

%However, {\name} still outperforms both methods in this dataset as it overcomes the issue in Sim-piece. In almost all the datasets, the proposed method consistently achieves a better compression ratio compared to both methods for the same epsilon values, indicating that {\name} is a more efficient piecewise approximation method than previous ones. We can also find that the impact of the error threshold on the compression ratio is significant. For smaller values of $\epsilon$, there is a steep decline in the compression ratio for all methods, especially APCA. The finding motivates us to select a medium error value to construct the base. We will talk about the effect of $\epsilon_{b}$ in ~\autoref{granularity}, where a better compression ratio can be achieved.

\subsection{Effect of base error threshold}\label{Sec:granularity}

%\gy{\autoref{fig:varerro} gives a more detailed comparison with APCA. Compared to \name, APCA is a more lightweight method, as it only keeps constant-value intervals without slopes. Therefore, it has an advantage at higher error thresholds, such as when $\epsilon=1e-1$. On the contrary, \name shows their advantage at lower error thresholds. The CR of APCA falls rapidly from a common value as the error falls, which also appears as growing rapidly as the error rises. However, \name can be configured to change the base quantization step and error threshold~$\epsilon_{b}$ to increase the compression ratio.} 

\begin{figure}[ht]
\vspace{-4mm}
\centering
\includegraphics[width=0.9\linewidth]{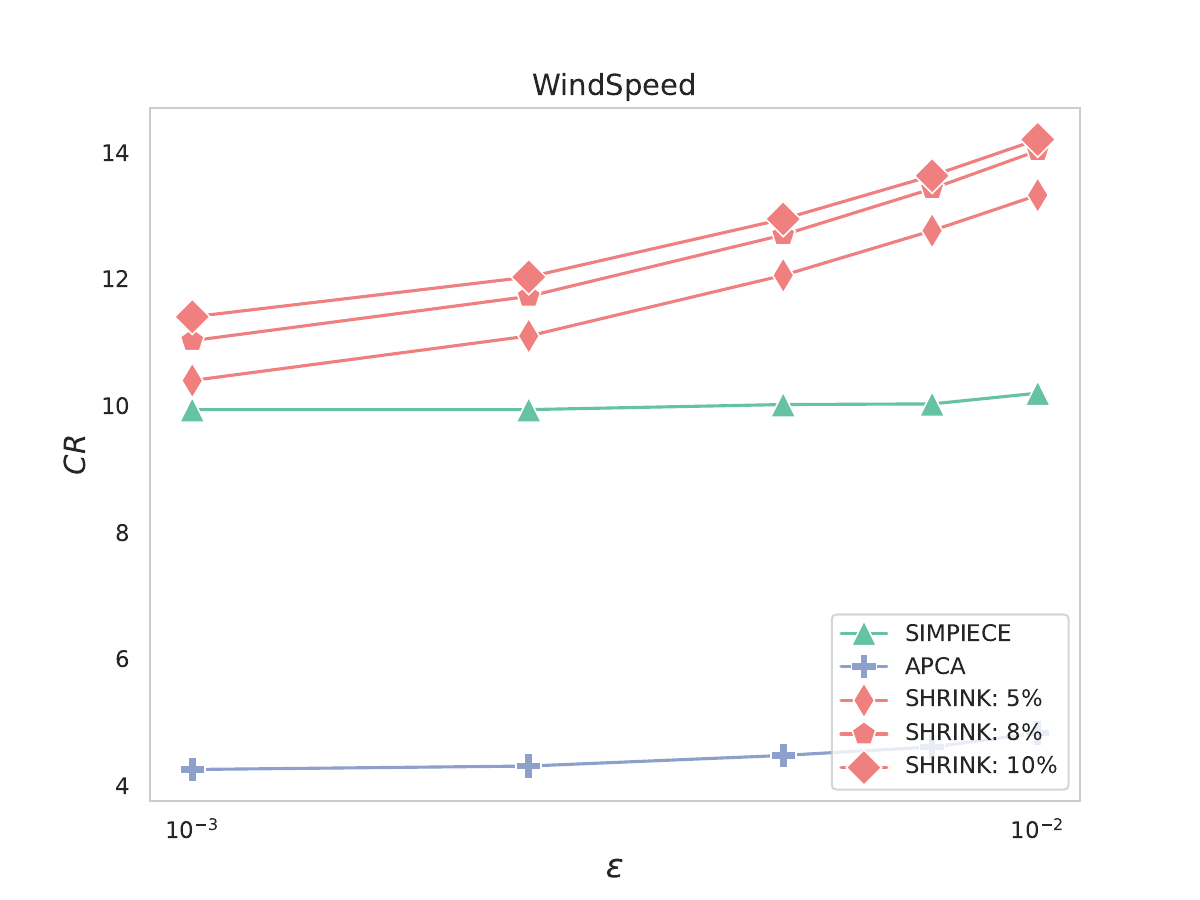}
\vspace{-4mm}
\caption{\pks{Effect of base error threshold $\epsilon_b$.}}\label{fig:varerro}
\vspace{-4mm}
\end{figure}

\pks{We now study how the compression ratio of \name depends on the base quantization step~$\epsilon_b$ that \name employs to define quantization invervals when extracting semantics to build its knowledge base. We use the WindSpeed data set and set~$\epsilon_b$ to~$5\%$, $8\%$ and~$10\%$ of the range of the dataset. \autoref{fig:varerro} shows our results. Notably, the compression ratio rises as we relax~$\epsilon_b$, since a larger~$\epsilon_b$ yields fewer cones, hence fewer sub-bases. While this effect also requires more residuals, the net effect is a reduction of the total data size. In a nutshell, the value of~$\epsilon_b$ trades off the size of base and residuals.}

\pks{For each dataset, it is in principle possible to find an optimal~$\epsilon_b$ that achieves the highest compression. Nevertheless, as we aim, apart from compression, to enable downstream analytics tasks using the semantic knowledge base without residuals, we suggest keeping~$\epsilon_b$ reasonably small, e.g., at~$5\%$ of the range. We thus opt for a base error threshold~$\epsilon_b$ that strikes a balance between compression ratio and accuracy in data analysis without residuals. On datasets with sharp discontinuities, where there is little meaningful semantics, we suggest relaxing~$\epsilon_b$. Besides, if a high compression ratio is the prime objective, a larger~$\epsilon_b$ may yield better results.}

\subsection{\pks{Effect of data set size}}\label{edge} %Enhancing edge server efficiency

To investigate how \name handles a growing data set size, we generate synthetic data of growing size by infusing noise drawn from a normal distribution~$\mathcal{N}(0, 0.1)$ to a classic scientific dataset, household power consumption data, reaching size above~1GB. We chose this data set because previous work~\cite{chandak2020lfzip} showed that linear-model-based compression methods performed poorly on it due to sharp discontinuities, which render approximation by piecewise linear functions or lower-order polynomials hard; we also observed this phenomenon with the WindSpeed and WindDirection data. We aim to test \name on this challenging data set.

\begin{table*}[!t]
\centering
% \hspace*{10cm}
\caption{\pks{Compression latency in sec, five lossless methods against {\ourname} with~$\epsilon \in \{0, 0.001, 0.01\}$.}}\label{tab:Time}
\vspace{-0.8em}
\begin{tabular}{|c|l|l|l|l|l|ccclll|ll}
\cline{1-12}
\multicolumn{1}{|l|}{\multirow{3}{*}{}} & \multirow{2}{*}{\textbf{Gzip}} & \multirow{2}{*}{\textbf{TRC}} & \multirow{2}{*}{\textbf{BZip2}} & \multirow{2}{*}{\textbf{Gorilla}} & \multirow{2}{*}{\textbf{GD}} & \multicolumn{6}{c|}{\textbf{\ourname}}                                                                                                                 &  &  \\ \cline{7-12}
\multicolumn{1}{|l|}{}                  &                                &                               &                                 &                                   &                              & \multicolumn{3}{c|}{\textbf{Base}}                                              & \multicolumn{3}{c|}{\textbf{Residual}}                                              &  &  \\ \cline{2-12}
\multicolumn{1}{|l|}{}                  & \multicolumn{1}{c|}{0}         & \multicolumn{1}{c|}{0}        & \multicolumn{1}{c|}{0}          & \multicolumn{1}{c|}{0}            & \multicolumn{1}{c|}{0}       & \multicolumn{1}{c|}{0} & \multicolumn{1}{c|}{0.001} & \multicolumn{1}{c|}{0.01} & \multicolumn{1}{c|}{0}     & \multicolumn{1}{c|}{0.001} & \multicolumn{1}{c|}{0.01} &  &  \\ \cline{1-12}
\textbf{FaceFour}                       & 0.09                           & 0.08                          & 0.04                            & 0.15                              & 0.41                         & \multicolumn{3}{c|}{0.07}                                                       & \multicolumn{1}{l|}{0.07}  & \multicolumn{1}{l|}{0.03}  & 0.03                      &  &  \\ \cline{1-12}
\textbf{MoteStrain}                     & 0.40                           & 0.17                          & 0.11                            & 0.53                              & 0.94                         & \multicolumn{3}{c|}{0.20}                                                       & \multicolumn{1}{l|}{0.18}  & \multicolumn{1}{l|}{0.09}  & 0.08                      &  &  \\ \cline{1-12}
\textbf{Lightning}                      & 0.35                           & 0.19                          & 0.12                            & 0.55                              & 1.28                         & \multicolumn{3}{c|}{0.17}                                                       & \multicolumn{1}{l|}{0.22}  & \multicolumn{1}{l|}{0.09}  & 0.08                      &  &  \\ \cline{1-12}
\textbf{ECG}                            & 1.69                           & 1.20                          & 0.73                            & 3.35                              & 5.46                         & \multicolumn{3}{c|}{1.25}                                                       & \multicolumn{1}{l|}{1.34}  & \multicolumn{1}{l|}{0.53}  & 0.44                      &  &  \\ \cline{1-12}
\textbf{Cricket}                        & 1.78                           & 1.26                          & 0.75                            & 3.44                              & 6.67                         & \multicolumn{3}{c|}{1.05}                                                       & \multicolumn{1}{l|}{1.25}  & \multicolumn{1}{l|}{0.52}  & 0.45                      &  &  \\ \cline{1-12}
\textbf{WindDirection}                  & 2.34                           & 1.56                          & 0.90                            & 5.57                              & 8.21                         & \multicolumn{3}{c|}{2.71}                                                       & \multicolumn{1}{l|}{1.35}  & \multicolumn{1}{l|}{1.06}  & 0.96                      &  &  \\ \cline{1-12}
\textbf{Wafer}                          & 2.07                           & 1.79                          & 1.02                            & 4.14                              & 12.74                        & \multicolumn{3}{c|}{1.93}                                                       & \multicolumn{1}{l|}{1.80}  & \multicolumn{1}{l|}{0.84}  & 0.67                      &  &  \\ \cline{1-12}
\textbf{WindSpeed}                      & 10.05                           & 5.47                          & 2.72                            & 19.52                             & 22.51                        & \multicolumn{3}{c|}{3.64}                                                      & \multicolumn{1}{l|}{3.56}  & \multicolumn{1}{l|}{3.43}  & 2.99                      &  &  \\ \cline{1-12}
\textbf{Pressure}                       & 38.37                          & 19.24                         & 9.13                            & 40.57                             & 62.59                        & \multicolumn{3}{c|}{4.36}                                                      & \multicolumn{1}{l|}{7.81} & \multicolumn{1}{l|}{5.61}  & 4.67                      &  &  \\ \cline{1-12}
\end{tabular}
\end{table*}

\begin{figure}[!h]
\vspace{-2mm}
\centering
\includegraphics[width=0.9\linewidth]{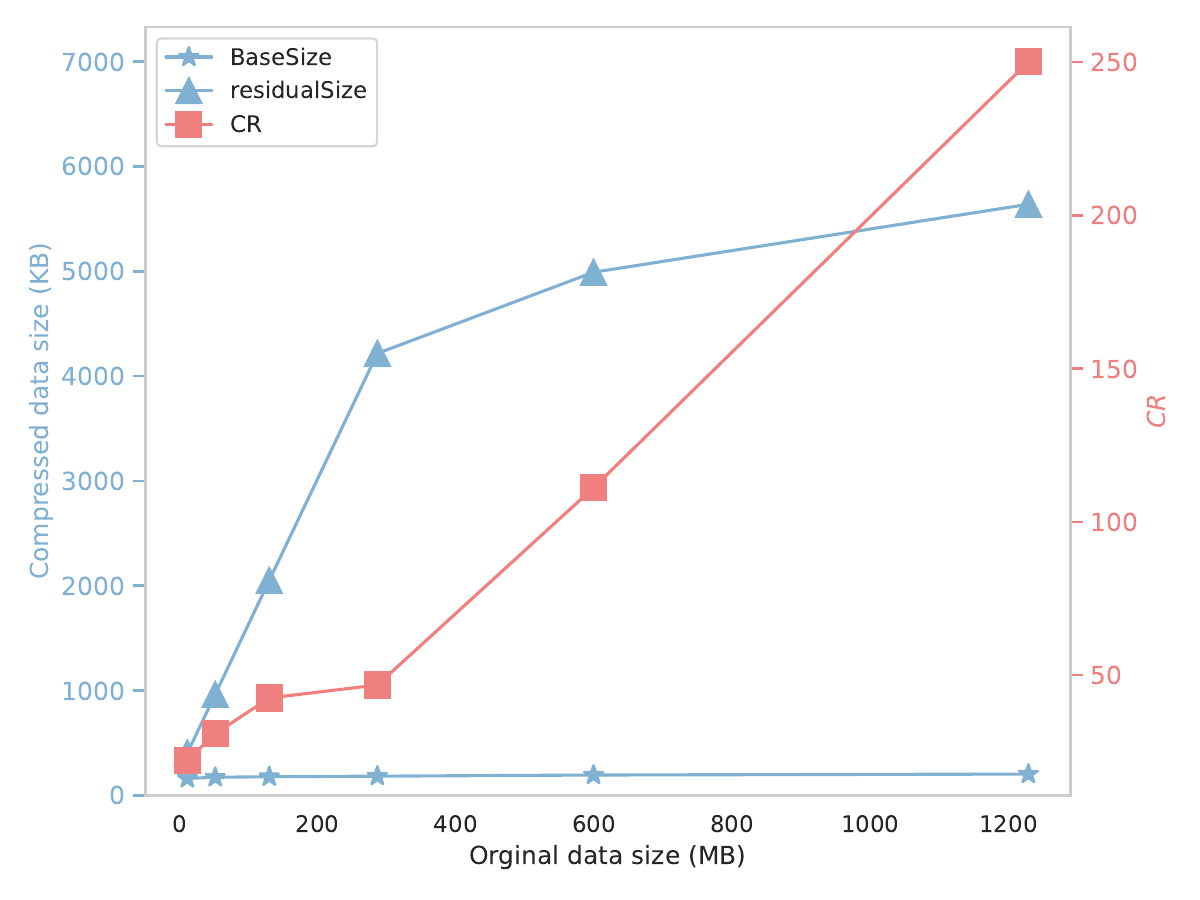}
\vspace{-5mm}
\caption{\pks{Effect of data size on the size of base and residuals.}}\label{fig:BaseChange}
%\name achieves high data retention under resource constrains.
\vspace{-2mm}
\end{figure}

\autoref{fig:BaseChange} depicts the dependence of base and residual sizes. Notably, the base remains relatively stable in size. By contrast, the residuals exhibit a linear growth. This steady growth is manageable and anticipated, as it aligns with the stochastic nature of noise and the ensuing necessity to capture novel information. Nevertheless, the marginal increase in the size of base amidst a considerable growth of the total dataset size testifies to the efficacy of \name in differentiating enduring data patterns from fluctuations. This capability is particularly beneficial on edge servers. By ensuring that only the essentials are stored, \name enables edge computing to overcome the storage limitations.

\subsection{Compression throughput}

%\autoref{tab:Time} illustrates the compression latency of . %\textcolor{purple}{It boasts significantly enhanced compression throughput compared with Sim-piece, especially in comparison to methods yielding lower compression ratios. This advantage is primarily due to {\name}'s ability to discern and encode most repeated data features at $\hat{\epsilon_{b}}$, whilst retaining the capability to adjust to dynamic resolutions. From this data, it is apparent that our method markedly outperforms Sim-piece, establishing itself as the most space-efficient and time-efficient algorithm.}

\begin{figure}[ht]
\vspace{-2mm}
\centering
\includegraphics[width=0.9\linewidth]{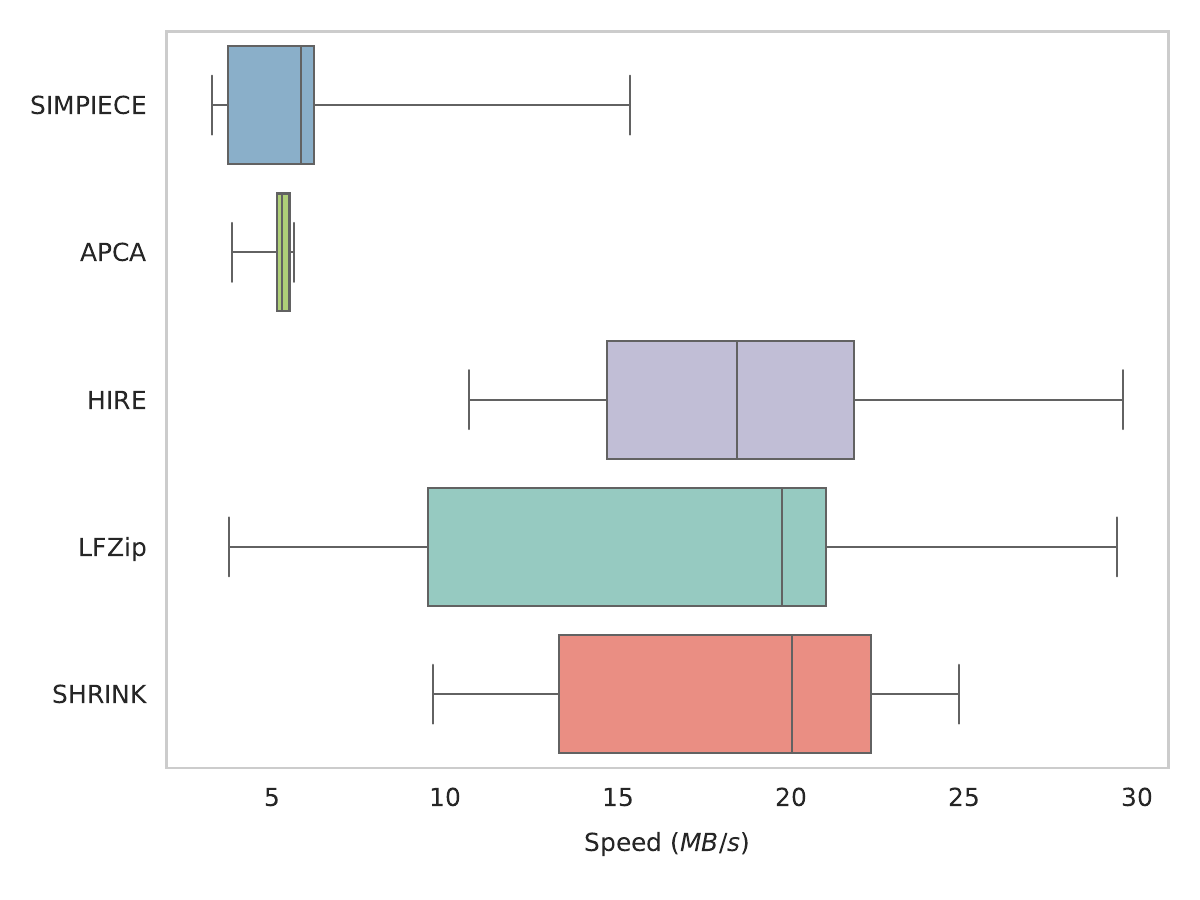}
\vspace{-6mm}
\caption{\pks{Compression throughput of five lossy methods.}}\label{fig:speed}
\vspace{-3mm}
\end{figure}

In this sub-section, we study the compression throughput of \name. Significantly, we implemented \name in Python and have not optimized it for time-efficiency yet. \pks{We compare \name to \simpiece, APCA, HIRE and~LFZip in terms of compression throughput.} \autoref{fig:speed} shows the distribution of throughputs for each compressor on the~9 datasets. To allow a fair comparison against \simpiece, we implement it in~Python too. We select ten different error thresholds~$\epsilon$ to compute the average throughput for each dataset. As can be seen, \name provides~3$\times$ speedup in compression in comparison with \simpiece and~APCA \pks{and} achieves comparable throughput compared to~HIRE and~LFZip. It is worth mentioning that~LFZip is written in~Python and~C++.
%Note that even higher compression throughput can attained with \name when there are more data requirements in edge by just applying compression on residuals.
We stress that, once the knowledge base is constructed in an operation that takes up most of the time, $\name$ only needs to encode residuals at different error resolutions. Thus, \name reduces the consumed time further as we already constructed base.

\pks{We further compare~$\name$ against the five lossless benchmarks. \autoref{tab:Time} lists the compression time for each dataset in seconds. We distinguish the time of \name into time for base construction and residual encoding and present it for three different error thresholds: $0$, $0.001$, and $0.01$.} Notably, in \name base construction takes up a significant portion of the total compression time, while the residual encoding is relatively fast. The main driver of \name’s performance advantage against others is that it uses a simple but effective PLA method to construct its base.

% \pks{Gorilla and GD perform considerably worse than other benchmarks. Gorilla employs a complex recursive encoding that compares adjacent values and executes bitwise operations, leading to low efficiency. GD, constructs a bit-level base, similarly to~\name, yet uses a complex greedy algorithm to find the optimal partition, hampering its efficiency. }

% This directly contrasts with the lossy benchmarks, each of which compresses the data once per error threshold to produce 10 separable encodings.
% We exclude HIRE and APCA from the comparison, since HIRE compresses a batch of data (with size of power of 2) rather than the entire dataset, and APCA has no advantage in compression ratio.
\subsection{Impact of default interval length}

\begin{figure}[hbt!]
\vspace{-3mm}
\centering
\includegraphics[width=0.9\linewidth]{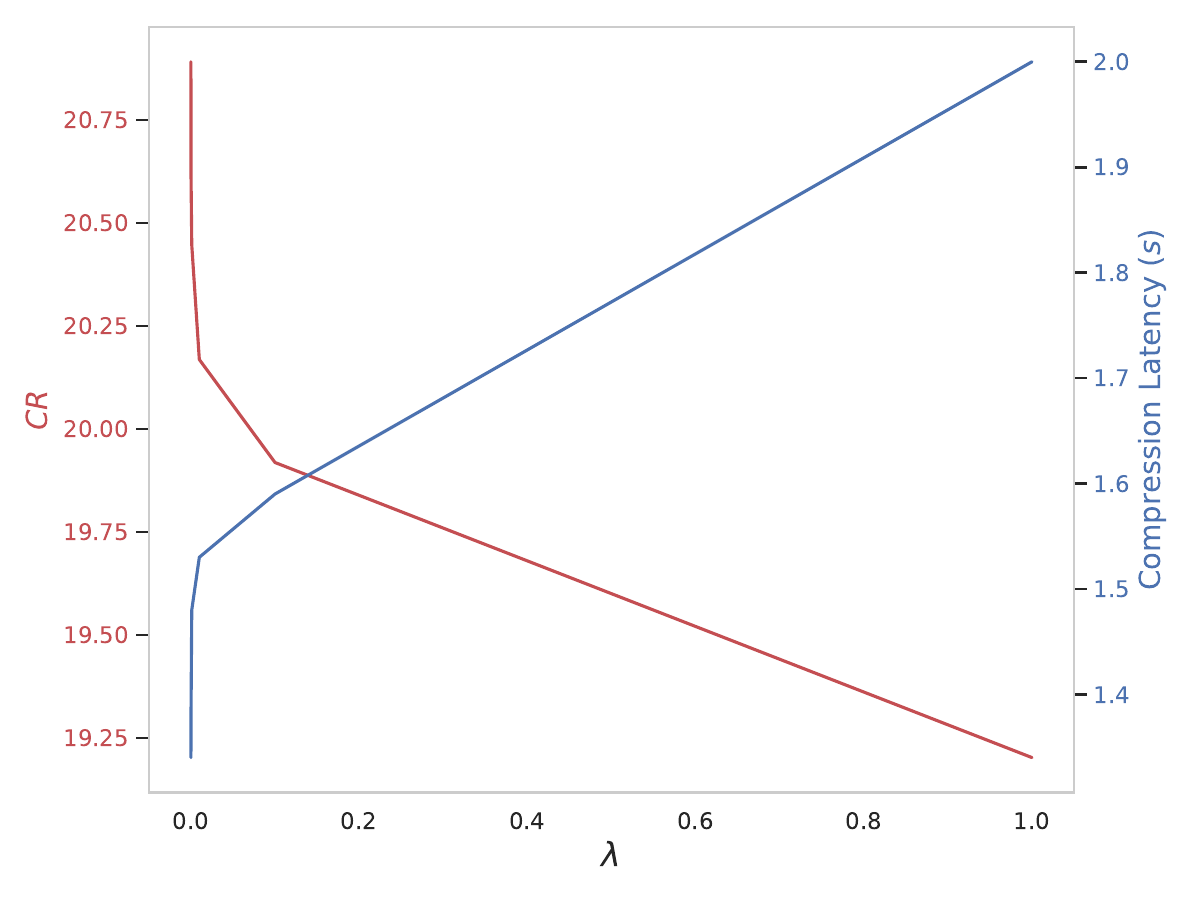}
\vspace{-3mm}
\caption{\pks{Compression ratio and latency vs.~$\lambda$.}}\label{fig:intervalsize}
\vspace{-2mm}
\end{figure}

As we have seen, an interval stores a subset of the entire dataset, to be used to extract semantics, affected by~$\epsilon_b$ and the parameter~$\lambda$ determining the default interval length. We now examine the effect of~$\lambda$ on the compression performance. The results in~\autoref{fig:intervalsize} show that, as~$\lambda$ falls, the compression ratio rises. This phenomenon can be explained by two reasons. On the one hand, a smaller~$\lambda$ results in a reduced interval length, which allows \name to identify the data's variance more thoroughly, decreasing the redundancy in semantic representation. On the other hand, smaller interval lengths confine the effects of outliers to lesser data portions. Consequently, a smaller default interval length reduces the volume of data retained, affecting the total compression ratio.

\autoref{fig:intervalsize} also portrays the effect of~$\lambda$ on compression latency. Notably, as~$\lambda$ grows, latency increases. Thus, the decrease in buffer size has a positive effect on compression latency. Starting from a buffer size where~$\lambda = 0.00001$, we witness a steep increase of compression latency as~$\lambda$ rises. Thereafter, compression latency changes less steeply. We attribute this fact to the lower data fluctuation in small-size buffers, which causes \name to increase its error bound when extracting semantics, hence a speedup.

As discussed, using an adaptive base error threshold $\hat{\epsilon}_b$ to extract semantics can preserve more features. Now we see that compression performance worsens as~$\lambda$ approaches~1. With~$\lambda = 1$, the whole dataset will be divided into two intervals, and~$\hat{\epsilon}_b$ is expected to be the same as the default~$\epsilon_b$, since the fluctuation level in so large intervals tends to be same as the whole dataset.
\section{Discussion}

\pks{Here we highlight the areas where \name performs most well, as well as its limitations.}

\noindent \textbf{General-purpose lossless methods} solely focus on compression performance and generally provide poor support for downstream applications. For instance, these methods necessitate to decompress the compressed data to retrieve even a single bit, hence are ill-suited for modern data storage systems. \name employs linear segment represent data points with the semantics and its compression performance is comparable to that of state-of-art general-purpose lossless methods. Moreover, the compression ratio of these general-purpose lossless methods does not improve much as the dataset size grows under similar patterns, whereas the compression ratio of \name increases with increasing dataset size in that case.

\noindent \textbf{\simpiece and other piecewise approximation methods} \pk{offer high compression performance, as they use a rather simple representation that encompasses many data points. Particularly, \simpiece captures similar patterns in time series data, and hence represents these data compactly to enhance compression ratio. However, compression performance degrades rapidly in the case of high-precision data recovery, e.g., $\epsilon=10^{-3}$, and becomes even worse than that of lossless compression. \name addresses this drawback and provides better compression performance for ultra-accurate data recovery.}

% \noindent \textbf{HIRE}, as a powerful multiresolution compression method, creates a single encoding that effectively yields multiple output resolutions. Its storage footprint can be rather small, as it keeps a single encoding with small compression and decompression latency. However, HIRE also suffers from performance degradation at high-precision data requests. According to our experimental results, the compression ratio of HIRE may even be below that of LFZip in some cases. \name also offers multiresolution compression, yet its focus is on ulta-accuracy multiresolution compression, where it achieves far better compression ratios than HIRE. Besides, HIRE only handles data batches with the size of a power of~2, not the whole dataset, rendering comparison to other methods hard. \name is more flexible, compressing data without such a limitation. 
\noindent \textbf{General-purpose lossy methods}, such as LFZip and HIRE, provide a stable compression ratio and high speed. Sometimes, their performance is even better than \simpiece, yet they do not provide sophisticated features, such as linear segment or random access. This deficiency limits their applicability to modern edge-based data infrastructure. Similarly, its compression ratio degrades rapidly with more strict precision requests.

% discussion in Section 4.4 (effect of data size) merged here.
\noindent \textbf{Scope and limitations of \name.} \pks{\name is commendable to enhance the use of storage by compressing large data sets with repeated patterns, especially in applications that need to recover high-precision historical data to perform analytical tasks on limited-storage equipment, such as Edge servers in the IoT ecosystem. However, \name pays less off on small datasets, as it has to extract semantics and construct a knowledge base first. Besides, its compression performance is less competitive when we do not need high precision, as with~$\epsilon \ge 10^{-1}$. Lastly, just like \simpiece and HIRE, \name does not natively support the multidimensional case (e.g., image compression), although it is extensible to multiple dimensions by encoding each column independently. We relegate the development of a multidimensional solution to future work.}

%\textcolor{purple}{Specifically, if distributed storage infrastructure with resource constraint exits, {\name} could provide rather impressive performance because we could make the local storage can be ignored. ??} One specific case in which {\name} is undesirable is when high compression throughput is needed because our methods aim at learning the pattern, which is relatively time-consuming compared to other general-purpose compression methods.
% \input{sections/06.related}
\section{Conclusion}

We introduced \name, a novel error-bounded data compression method based on semantic extraction and residual encoding. Compared to prior works, \name drastically improves compression at comparable speeds and avoids degrading compression performance when aiming for ultra-accurate data recovery. \name extracts piecewise linear segments in a first, data-level compression phase while \emph{adapting} its error tolerance to data fluctuations, thereby detecting data patterns that it uses to construct its knowledge base; further, it merges recurrent similar linear-segment patterns to achieve further compression. In a second, bit-level compression phase, \name encodes the \emph{residuals} subtracted from the base. Our thorough experiments demonstrate that \name outperforms state-of-art lossless and lossy compressors.

% \pks{We introduced \name, a novel error-bounded data compression method based on semantic extraction and residual encoding. Compared to prior works, \name drastically improves compression at comparable speeds and avoids degrading compression performance when aiming for ultra-accurate data recovery. The core novelty of \name is that it extracts piecewise linear segments in a first, data-level compression phase while \emph{adapting} its error tolerance to data fluctuations, thereby detecting data patterns that it uses to construct its knowledge base; further, it merges recurrent similar linear-segment patterns to achieve further compression. In a second, bit-level compression phase, \name encodes the \emph{residuals} subtracted from the base. Besides, \name offers multi-resolution compression and addresses the drawbacks of previous multi-resolution methods in the domain of highly accurate data recovery. Our thorough experiments demonstrate that \name outperforms state-of-art lossless and lossy compressors.}
\section*{Acknowledgment}
This work is supported by Independent Research Fund Denmark Light-IoT project \emph{Analytics Straight on Compressed IoT Data} (Grant No. 0136-00376B),  Innovation Fund Denmark GreenCOM project (Grant No. 2079-00040B), NordForsk Nordic University Cooperation on Edge Intelligence (Grant No. 168043) and Aarhus University DIGIT Centre.

\bibliography{sample}

\end{document}